\documentclass[aps,preprint,nofootinbib,a4paper,11pt]{revtex4-1}
\usepackage{amsmath} 
\usepackage{graphicx} 
\usepackage{amsthm}
\usepackage{amssymb} 
\usepackage{dsfont}
\usepackage{yfonts}
\usepackage{hyperref}
\usepackage{floatrow,dcolumn,rotating}
\usepackage{array,xcolor,graphicx}
\usepackage{booktabs,multirow}
\usepackage[utf8]{inputenc}

\hypersetup{colorlinks=true,linkcolor=blue,citecolor=red,urlcolor=blue}

\newcommand{\be}{\begin{equation}}
\newcommand{\ee}{\end{equation}}
\newcommand{\bea}{\begin{eqnarray}}
\newcommand{\eea}{\end{eqnarray}}

\newcommand{\nn}{\nonumber\\}

\def\la{\langle}
\def\ra{\rangle}
\def\Tr{{\mathrm{Tr}}}
\def\tr{{\mathrm{tr}}}

\def\d{\partial}
\def\grad{\nabla}
\def\CC{\mathcal{C}}

\def\CF{\mathcal{F}}
\def\CG{\mathcal{G}}
\def\CH{\mathcal{H}}
\def\CI{\mathcal{I}}
\def\CJ{\mathcal{J}}

\def\CL{\mathcal{L}}

\def\CN{\mathcal{N}}
\def\CS{\mathcal{S}}

\def\CO{\mathcal{O}}
\def\CP{\mathcal{P}}

\begin{document}

\title{Universality of anomalous conductivities in theories \\with higher-derivative holographic duals}
\author{S.~Grozdanov}
\email{grozdanov@lorentz.leidenuniv.nl}
\author{N.~Poovuttikul}
\email{poovuttikul@lorentz.leidenuniv.nl}
\affiliation{Instituut-Lorentz for Theoretical Physics, \\Leiden University, Niels Bohrweg 2, \\Leiden 2333 CA, The Netherlands\\
\vspace{1cm}
}
\begin{abstract}
\vspace{0.5cm}
Anomalous chiral conductivities in theories with global anomalies are independent of whether they are computed in a weakly coupled quantum (or thermal) field theory, hydrodynamics, or at infinite coupling from holography. While the presence of dynamical gauge fields and mixed, gauge-global anomalies can destroy this universality, in their absence, the non-renormalisation of anomalous Ward identities is expected to be obeyed at all intermediate coupling strengths. In holography, bulk theories with higher-derivative corrections incorporate coupling constant corrections to the boundary theory observables in an expansion around infinite coupling. In this work, we investigate the coupling constant dependence and universality of anomalous conductivities (and thus of the anomalous Ward identities) in general, four-dimensional systems that possess asymptotically anti-de Sitter holographic duals with a non-extremal black brane in five dimensions, and anomalous transport introduced into the boundary theory via the bulk Chern-Simons action. We show that in bulk theories with arbitrary gauge- and diffeomorphism-invariant higher-derivative actions, anomalous conductivities, which can incorporate an infinite series of (inverse) coupling constant corrections, remain universal. Owing to the existence of the membrane paradigm, the proof reduces to a construction of bulk effective theories at the horizon and the boundary. It only requires us to impose the condition of horizon regularity and correct boundary conditions on the fields. We also discuss ways to violate the universality by violating conditions for the validity of the membrane paradigm, in particular, by adding mass to the vector fields (a case with a mixed, gauge-global anomaly) and in bulk geometries with a naked singularity.
\end{abstract}

\maketitle
\begingroup
\hypersetup{linkcolor=black}
\tableofcontents
\endgroup

\section{Introduction}\label{sec:Intro}
{\bf Anomalies}

An anomaly is a quantum effect whereby a classically conserved current $J^\mu$ ceases to enjoy its conservation, $\nabla_\mu \la J^\mu \ra \neq 0 $ \cite{Weinberg:1996kr,Bertlmann:1996xk,Bilal:2008qx,Harvey:2005it}. To date, a multitude of different anomalies have been discovered that can be classified into two main categories: local (gauge) and global anomalies. A gauge anomaly corresponds to a gauged symmetry (and current) and the consistency of a quantum field theory requires this anomaly to vanish. While global anomalies are permitted, their existence still imposes stringent conditions on the structure of quantum field theories due to the anomaly matching condition discovered by 't Hooft \cite{'tHooft:1979bh}. The condition states that a result of an anomaly calculation must be invariant under the renormalisation group flow and is thus independent of whether it is computed in the UV microscopic theory or an IR effective theory. 

Of particular importance to quantum field theory have been the chiral anomalies, which are present in theories with massless fermions. The values of the current divergences resulting from these anomalies are known to be one-loop exact. From the point of view of the topological structure of gauge theories, one can suspect that this should be true very generically due to the fact that the anomaly is related to the topologically protected index of the Dirac operator. Perturbatively, non-renormalisation of the one-loop anomalies was established in \cite{Adler:1969gk,Bell:1969ts,PhysRev.182.1517}. In a typical four dimensional chiral theory, there are two classically conserved currents: the axial $J^\mu_5$ (associated with the $\gamma_5$ Dirac matrix) and the vector current $J^\mu$. By including quantum corrections, their Ward identities can be written as
\begin{equation}
\begin{aligned}
\nabla_\mu \left\langle J^\mu_5 \right\rangle &= \epsilon^{\mu\nu\rho\sigma}\left(\kappa
  {F}_{A,\mu\nu} {F}_{A,\rho\sigma} + \gamma {F}_{V,\mu\nu}
  {F}_{V,\rho\sigma} + \lambda {R}^{\alpha_1}_{\;\;\alpha_2 \mu\nu}
  {R}^{\alpha_2}_{\;\;\alpha_1 \rho\sigma}   \right),\\
\nabla_\mu \left\langle J^\mu \right\rangle &= 0,
\end{aligned}
\label{eqn:non-conserved-current}
\end{equation}
where $F_{A,\mu\nu}$, $F_{V,\mu\nu}$ are the field strengths associated with the axial and the vector gauge fields. $R^{\alpha}_{\;\;\beta\mu\nu}$ is the Riemann curvature tensor of the curved manifold on which the four dimensional field theory is defined, and $\kappa$, $\gamma$ and $\lambda$ are the three Chern-Simons coupling constants. While the axial current conservation is violated by quantum effects, the vector current remains conserved. Among other works, various arguments in favour of non-renormalisation of one-loop anomalies have been presented in \cite{Fukushima:2008xe,Newman:2005as,Kharzeev:2009pj,Jensen:2012jy,Banerjee:2012iz,Nair:2011mk,Sadofyev:2010pr,Sadofyev:2010is}. The situation is much less clear when, as in \cite{Jensen:2013vta}, one considers the contributions of mixed, gauge-global anomalies. In such cases, it was shown in \cite{Jensen:2013vta} that one  should expect anomalous currents to receive radiative corrections at higher loops. The connection between this work and mixed, gauge-global anomalies will be elaborated upon below. A further set of open questions related to the non-renormalisation of anomalies enters the stage from the possibility of considering non-perturbative effects in QFT.

From a historically more unconventional point of view, anomalies have recently also been studied through the (macroscopic) hydrodynamic entropy current analysis \cite{Son:2009tf,Neiman:2010zi}.\footnote{For a recent discussion of anomalies from the point of view of UV divergences in classical physics and its connection to the breakdown of the time reversal symmetry, see \cite{Polonyi:2013lma,Polonyi:2015tla}.} The effects of gravitational anomalies on the hydrodynamic gradient expansion were then studied by using the Euclidean partition function on a cone in \cite{Jensen:2012kj}. Macroscopic transport properties associated with anomalous conservation laws have now been analysed in detail (at least theoretically) both at non-zero temperature and density. To date, the most prominent and well-understood anomaly-induced transport phenomena have been associated with the chiral magnetic effect \cite{Kharzeev:2007tn,Fukushima:2008xe,Kharzeev:2009pj} and the chiral vortical effect \cite{Son:2009tf,Kharzeev:2010gr}. 
\\

{\bf Chiral conductivities in field theory}

In the low-energy hydrodynamic limit, we expect that to leading order in the gradient expansion of relevant fields, the expectation values of these currents can be expressed in the form of Ohm's law. The corresponding conductivities can then be defined in the following way: If a chiral system is perturbed by a small external magnetic field $B^\mu = (1/2)\epsilon^{\mu\nu\rho\sigma} u_\nu F_{\rho\sigma}$ and a spacetime vortex $\omega^\mu = \epsilon^{\mu\nu\rho\sigma}u_\nu \nabla_\rho u_\sigma$, where $u^\mu$ is the fluid velocity vector in the laboratory frame, then the expectation values of the two currents change by $\la \delta J^\mu\ra$ and $\la \delta J^\mu_5\ra$. Note that unlike in Eq. \eqref{eqn:non-conserved-current}, both the axial and vector current conservation are now broken by the induced anomalies. To leading (dissipationless) order, the change can be expressed in terms of the conductivity matrix
\begin{align}\label{CondDef}
\left(      
\begin{array}{c}
\la \delta J^\mu \ra \\
\la \delta J^\mu_5 \ra
\end{array}
\right) = 
\left(
\begin{array}{cc}
\sigma_{JB} & \sigma_{J\omega} \\
\sigma_{J_5 B} & \sigma_{J_5\omega}
\end{array}
\right)
\left(
\begin{array}{c}
B^\mu \\
\omega^\mu
\end{array}
\right),
\end{align}
where $\sigma_{JB}$ is known as the {\em chiral magnetic conductivity}, $\sigma_{J\omega}$ as the {\em chiral vortical conductivity} and $\sigma_{J_5 B}$ as the {\em chiral separation conductivity}. The signature of anomalies can thus be traced all the way to the extreme IR physics and analysed by the linear response theory. This will be the subject studied in this work.

By following a set of rules postulated in \cite{Loganayagam:2012zg} (see also \cite{Jensen:2013rga}), a convenient way to express the anomalous conductivities is in terms of the anomaly polynomials. We briefly review these rules in Appendix \ref{app:anomP}. They allow one to compute the anomalous conductivities from the structure of the anomaly polynomials in arbitrary (even) dimensions, independently of the value of the coupling constant \cite{Jensen:2012kj,Azeyanagi:2013xea,Loganayagam:2012zg,Jensen:2013rga}.

In the IR limit, we may assume that the stress-energy tensor and the charge current can
be expressed in a hydrodynamic gradient expansion
\cite{Kovtun:2012rj,Baier:2007ix,Romatschke:2009kr,Grozdanov:2015kqa}. The constitutive
relations for a fluid with broken parity, in the Landau frame, are \cite{Erdmenger:2008rm,Son:2009tf,Banerjee:2008th,Torabian:2009qk}
\begin{equation}
\begin{aligned}
T^{\mu\nu}&= \varepsilon u^\mu u^\nu + P \Delta^{\mu\nu}  - \eta
\sigma^{\mu\nu} - \zeta \Delta^{\mu\nu} \grad_\lambda u^\lambda+ \CO\left(\d^2\right),\\
J^\mu_{I} &= n_{I} u^\mu + \sigma_{I} \Delta^{\mu\nu} \left( u^\rho F_{I,\rho\nu} - T\, \grad_\nu
  \left( \frac{\mu_{I}}{T}\right)\right) +\xi_{I,B} B_{I}^\mu + \xi_{I,\omega} \omega^\mu + \CO\left(\d^2\right),
\end{aligned}
\end{equation}
where the index $I=\{A,V\}$ labels the axial and the vector currents ($J^\mu_5 = J^\mu_A$, $J^\mu = J^\mu_V$) and their respective transport coefficients. In the stress-energy tensor, $\varepsilon$, $P$, $\eta$ and $\zeta$
are the energy density, pressure, shear viscosity and bulk viscosity. Furthermore, $n$,
$\sigma$, $T$, $\mu$ and $F_{\mu\nu}$ are the charge density, charge conductivity, temperature, chemical potential and the gauge field strength tensor. The vector field $u^\mu$ is the velocity field of the fluid, the transverse projector (to the fluid
flow) $\Delta^{\mu\nu}$ is defined as $\Delta^{\mu\nu} = u^\mu u^\nu + g^{\mu\nu}$, with $g^{\mu\nu}$ the metric tensor and $\sigma^{\mu\nu}$ the symmetric, transverse and traceless relativistic shear tensor composed of $\nabla_\mu u_\nu$. Plugging the above constitutive relations into the anomalous Ward identities, one can show that the anomalous conductivities are controlled by the transport coefficients $\xi_B$ and $\xi_\omega$ (see e.g. \cite{Landsteiner:2011iq}). It was shown in \cite{Son:2009tf,Neiman:2010zi} that by demanding the non-negativity of local entropy production (and similarly, by using a Euclidean effective action in \cite{Jensen:2012kj,Jensen:2012jh,Banerjee:2012iz})\footnote{Note that the analysis in \cite{Son:2009tf,Jensen:2012jh,Banerjee:2012iz} only involves the axial gauge field. However, it
is straightforward to generalise their results to the case with both the axial and the vector current.}, the anomalous chiral separation conductivity $\sigma_{J_5B}$ and the chiral magnetic conductivity $\sigma_{JB}$ become fixed by the anomaly coefficient $\gamma$:
\begin{align}\label{s1}
\sigma_{J_5 B} = -2 \gamma \mu, && \sigma_{J B} = -2 \gamma \mu_5.
\end{align}
On the other hand, the transport coefficient $\sigma_{J_5\omega} $ could not be completely determined by the anomaly and thermodynamic quantities. Its form contains an additional constant term, 
\begin{align}\label{s2}
\sigma_{J_5\omega} = \kappa \mu^2 + \tilde{c} T^2, 
\end{align}
where $\tilde{c}$ is some yet-undetermined constant, which could run along the renormalisation group flow. By using perturbative field theory methods \cite{Landsteiner:2011cp,Loganayagam:2012pz} and simple holographic models \cite{Landsteiner:2011iq,Azeyanagi:2013xea}, it was then suggested that $\tilde{c}$ could be fixed by the gravitational anomaly coefficients, $\lambda$.\footnote{We note that in the presence of chiral gravitinos, the relation between $\tilde c$ and the gravitational anomaly coefficient $\lambda$ is different from those studied in this work \cite{Loganayagam:2012pz,Chowdhury:2015pba}.} However, the gravitational anomaly enters the equations of motion \eqref{eqn:non-conserved-current} with terms at fourth order in the derivative expansion while $\xi_\omega$ and $\xi_B$ enter the equation of motion at second order. Thus, if one analysed the hydrodynamic expansion in terms of the na\"{i}ve gradient expansion with all fluctuations of the same order, it would seem to be impossible to express  $\tilde{c}$ in terms of the gravitational anomaly. The above paradox was resolved in \cite{Jensen:2012kj}. There, the theory was placed on a product space of a cone and a two dimensional manifold. The deficit angle $\delta$ was defined along the thermal cycle, $\beta$, as $\beta \sim \beta + 2\pi (1+\delta) $. Demanding continuity of one-point functions in the vicinity of $\delta = 0$ then fixed the unknown coefficient $\tilde{c}$ in terms of the gravitational anomaly coefficient $\lambda$ (the gradient expansion breaks down). The above construction can be extended to theories outside the hydrodynamic regime in arbitrary even dimensions and in the presence of other types of anomalies, so long as the theories only involve background gauge fields and a background metric \cite{Jensen:2013rga}. 

In the presence of dynamical gauge fields, the anomalous transport coefficients do not seem to remain protected from radiative corrections. This is consistent with the fact that the chiral vortical conductivity $\sigma_{J\omega}$, given otherwise by the thermal field theory result
\begin{align}\label{s3}
\sigma_{J \omega} = 2\gamma \mu_5 \mu,
\end{align} 
was also argued to get renormalised in theories with dynamical gauge fields by \cite{Golkar:2012kb,Hou:2012xg,Gorbar:2013upa}.\footnote{For a discussion of temperature dependence and thermal corrections to the chiral vortical conductivity in more complicated systems, see Ref. \cite{Kalaydzhyan:2014bfa}.} Furthermore, these various pieces of information regarding the renormalisation of the chiral conductivities are consistent with the findings of \cite{Jensen:2013vta} (already noted above) and lattice results \cite{Huang:1989kg,Yamamoto:2011ks,Yamamoto:2011gk,Fukushima:2010zza}: In theories with dynamical gauge fields and mixed, gauge-global anomalies, chiral conductivities renormalise.  
\\

{\bf Holography and universality of transport coefficients}

\begin{figure}[t]
\centering
\includegraphics[width=1\textwidth]{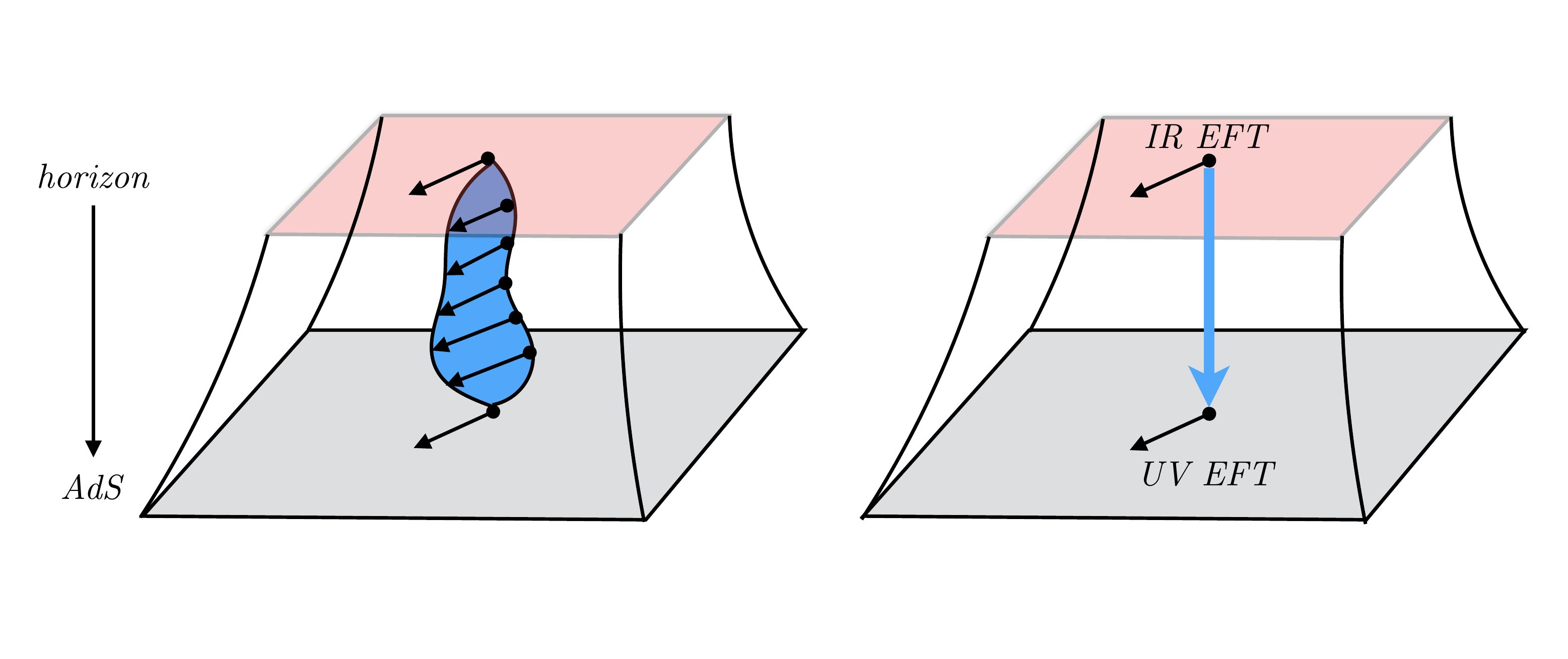}
\caption{A schematic representation of the membrane paradigm: The image on the left-hand-side corresponds to a holographic calculation (without the membrane paradigm) in which one has to solve for the bulk fields all along the $D$ dimensional bulk. On the right-hand-side (the membrane paradigm case), the field theory observable of interest can be read off from a conserved current (along the radial coordinate). Hence, we only need information about its dynamics at the horizon and the AdS boundary. The membrane paradigm enables us to consider independent {\em effective theories} at the two surfaces with $(D-1)$ dimensions. While the UV effective theory directly sources the dual field theory, it is the IR theory on the horizon that fixes the values of dual correlators in terms of the bulk black hole parameters. As in this paper, such a structure may enable us to make much more general (universal) claims about field theory observables then if the calculation depended on the details of the full $D$--dimensional dynamics.}
\label{fig_membranes}
\end{figure}

Certain classes of strongly interacting theories at finite temperature and chemical potential can be formulated using gauge-gravity (holographic) duality. Thus, in comparison with the weakly coupled regime accessible to perturbative field theory calculations, holography can be seen as a convenient tool to investigate chiral transport properties at the opposite end of the coupling constant scale. Within holography, anomalous hydrodynamic transport was first studied in the context of fluid-gravity correspondence \cite{Bhattacharyya:2008jc} by \cite{Kalaydzhyan:2011vx,Erdmenger:2008rm,Banerjee:2008th} who added the Chern-Simons gauge field to the bulk. The two DC conductivities associated specifically with chiral magnetic and chiral vortical effects were then computed in the five-dimensional anti-de Sitter Reissner-N\" ordstrom black brane background in \cite{Gynther:2010ed,Amado:2011zx,Landsteiner:2011iq}. The results were extended to arbitrary dimensions in \cite{Azeyanagi:2013xea}. The work of \cite{Azeyanagi:2013xea} showed that these transport coefficients could be extracted from first-order differential equations (as opposed to the usual second-order wave equations in the bulk) due to the existence of a {\em conserved current} along the holographic radial direction. In a similar manner, this occurs in computations of the shear viscosity \cite{Policastro:2001yc,Kovtun:2004de} and other DC conductivities \cite{Iqbal:2008by,Donos:2015gia}. We will refer to this situation as the case when the membrane paradigm is applicable (see Fig. \ref{fig_membranes}). The existence of the membrane paradigm makes the calculation of chiral conductivities significantly simpler. Reassuringly, the holographic results for the chiral conductivities agree with the results obtained from conventional QFT methods described above and stated in Eqs. \eqref{s1}, \eqref{s2} and \eqref{s3} \cite{Landsteiner:2011cp,Loganayagam:2012zg,Loganayagam:2012pz}. More recently, these calculations were generalised to cases of non-conformal holography (in which $T^\mu_{\;\;\mu} \neq 0$), giving the same results \cite{Gursoy:2014ela,Gursoy:2014boa}. A way to think of such holographic setups is as of geometric realisations of the renormalisation group flows. 

Universal holographic statements, most prominent among them being the ratio of shear viscosity to entropy density, $\eta / s = \hbar / (4 \pi k_B)$ \cite{Policastro:2001yc,Kovtun:2004de,Iqbal:2008by}, can normally be reduced to an analysis of the dynamics of a minimally-coupled massless scalar mode and the existence of the membrane paradigm. The fact that the membrane paradigm exists in some theories for anomalous chiral conductivities thus naturally leads to the possibility of universality of these transport coefficients in holography. Motivated by this fact, in this work, we study whether and when non-renormalisation theorems for anomalous transport can be established in holography.

Recently, a work by G\"{u}rsoy and Tarr\' io \cite{Gursoy:2014boa} made the first step in this direction by proving the universality of chiral magnetic conductivity $\sigma_{JB}$ in a two-derivative Einstein-Maxwell-dilaton theory with an arbitrary scalar field potential and anomaly-inducing Chern-Simons terms. The only necessary assumptions were that the bulk geometry is asymptotically anti-de Sitter (AdS) and that the Ricci scalar at the horizon must be regular. Because this statement is valid for two-derivative theories, it applies to duals at infinitely strong ('t Hooft) coupling $\lambda$ and infinite number of adjoint colours, $N$. In this sense, it is applicable within the same class of theories as the statement of universality for $\eta / s$.

Higher-derivative corrections to supergravity actions arise when $\alpha'$ corrections are computed from string theory. Usually, this is done by either computing loop corrections to the $\beta$-functions of the sigma model or by computing string scattering amplitudes and guessing the effective supergravity action that could result in the same amplitudes (see e.g. \cite{Grisaru:1986px,Gross:1986iv,Gross:1986mw}). Via the holographic dictionary, these higher-derivative corrections translate into (perturbative) coupling constant corrections in powers of the inverse coupling constant ($1/\lambda$) expanded around $\lambda \to \infty$ \cite{Gubser:1998nz}. The result of $ \eta / s = 1 / (4\pi)$ (having set $\hbar = k_B = 1$) is not protected from higher-derivative bulk corrections; it receives coupling constant corrections both in four-derivative theories (curvature-squared) \cite{Kats:2007mq,Brigante:2007nu,Buchel:2008wy,Myers:2009ij} and in the presence of the leading-order top-down corrections to the $\CN = 4$ supersymmetric Yang-Mills theory with an infinite number of colours (these $R^4$ corrections are proportional to $\alpha'^3 \sim 1 / \lambda^{3/2}$) \cite{Buchel:2004di}. An equivalent statement exists also in second-order hydrodynamics \cite{Baier:2007ix,Bhattacharyya:2008jc}. There, a particular linear combination of three transport coefficients, $2 \eta \tau_\Pi - 4 \lambda_1 - \lambda_2$, was shown to vanish for the same class of two-derivative theories as those that exhibit universality of $\eta /s $. It was then shown that the same linear combination of second-order transport coefficient vanishes to leading order in the coupling constant corrections even when curvature-squared terms \cite{Shaverin:2012kv,Grozdanov:2014kva} and $R^4$ terms dual to the $\CN=4$ 't Hooft coupling corrections are included in the bulk action \cite{Grozdanov:2014kva}. However, by using the non-perturbative results for these transport coefficients in Gauss-Bonnet theory \cite{Grozdanov:2015asa}, one finds that the universal relation is violated non-perturbatively (or at second order in the perturbative coupling constant expansion) \cite{Grozdanov:2014kva}.\footnote{The violation of universality in second-order hydrodynamics was later also verified in \cite{Shaverin:2015vda} by using fluid-gravity methods in Gauss-Bonnet theory.} 

Our goal in this work is to study the universality of the four anomalous conductivities $\sigma_{JB}$, $\sigma_{J\omega}$, $\sigma_{J_5 B}$ and $\sigma_{J_5\omega}$ in general higher-derivative theories, thereby incorporating an infinite series of coupling
constant corrections to results at infinite coupling (from two-derivative bulk theories). What we will show is that the expressions \eqref{s1}, \eqref{s2} and \eqref{s3} remain universal in any higher-derivative theory so long as the action (excluding the Chern-Simons terms) is gauge- and diffeomorphism-invariant.\footnote{As we are mainly interested in theories in which the anomalous Ward identity retains the form of Eq. \eqref{eqn:non-conserved-current}, the conditions of gauge- and diffeomorphism-invariance are imposed to avoid explicit violation of Eq. \eqref{eqn:non-conserved-current} by the bulk matter content (see Section \ref{section:massive-vector-fields} for a discussion of such an example that includes massive vector fields).} All we will assume, in analogy with \cite{Gursoy:2014boa}, is that the bulk theory is asymptotically AdS (it has a UV conformal fixed point) and that it permits a black brane solution with a regular, non-extremal horizon. In its essence, the proof will reduce to showing the validity of the membrane paradigm and then a study of generic, higher-derivative effective theories (all possible terms present in the conserved current) at the horizon and the boundary (as depicted in Fig. \ref{fig_membranes}). The condition of regularity of these constructions at the horizon will play a crucial role in the proof. By studying cases of theories for which the membrane paradigm fails, one can then find theories in which universality may be violated. 

Our findings can be seen as a test of holography in reproducing the correct Ward identities for the anomalous currents. The fact that we find universality of chiral conductivities with an infinite series of coupling constant corrections (albeit expanded around infinite coupling) is an embodiment of the fact that when only global anomalies are present, anomalous transport is protected from radiative corrections. An example related to the presence of mixed, gauge-global anomalies, which will invalidate the membrane paradigm,  will be studied in Section \ref{section:examples}. Again, as expected from field theory arguments, a case like that will naturally be able to violate the universality (or non-renormalisation) of chiral conductivities.  
\\

{\bf Organisation of the paper}

The paper is organised as follows: In Section \ref{section:holographic setup}, we describe the holographic theory at finite temperature and chemical potential that is studied in the main part of this work. We then turn to the proof of the universality of chiral conductivities in Section \ref{section:anomaly-induced-electric-current}. First, in Section \ref{sec.MemParad}, we show how to compute anomalous conductivities by using the membrane paradigm and specify the conditions that must be obeyed in order for the membrane paradigm to be valid. In Section \ref{section:any-higher-derivative}, we then prove that a gauge- and diffeomorphism-invariant action indeed satisfies those conditions and thus always gives the same anomalous conductivities. In Section \ref{section:examples}, we study examples that obey and violate the conditions required for universality. In particular, those that violate the universality include either massive gauge fields or naked singularities in the bulk. The paper proceeds with a discussion of results and future directions in Section \ref{section:discussion}. Finally, Appendix \ref{app:anomP} includes a discussion of anomaly polynomials and the replacement rule. 

\section{The holographic setup}
\label{section:holographic setup}

In this work, we consider five dimensional bulk actions with a dynamical metric $G_{ab}$, two massless gauge fields $A_a$ and $V_a$ that are dual to the axial and the vector current in the boundary theory, respectively, and a set of scalar (dilaton) fields, $\phi_I$: 
\begin{align}\label{FullAction}
S =  \int d^5x \sqrt{-G}\,  \left\{ \CL \left[A_a,V_a,G_{ab},\phi_I \right] +  \CL_{CS} \left[A_a,V_a,G_{ab} \right]  \right\}.
\end{align}
The Lagrangian density $\CL$ should be thought of as a general, diffeomorphism- and gauge-invariant action that may include arbitrary higher-derivative terms of the fields. Since we are interested in anomalous transport, \eqref{FullAction} must include the Chern-Simons terms, $\CL_{CS}$, that source global chiral anomalies in the boundary theory. In holography, higher-than-second-derivative bulk terms correspond to the ('t Hooft) coupling corrections to otherwise infinitely strongly coupled states ($\lambda \to \infty$). Since $\CL$ may include operators with arbitrary orders of derivatives (and corresponding bulk coupling constants), holographically computed quantities describing a hypothetical dual of \eqref{FullAction} are able to incorporate an infinite series of coupling constant corrections to observables at infinite coupling.\footnote{In type IIB theory, higher-derivative bulk terms and corrections to infinitely coupled results in $\CN = 4$ theory are proportional to powers of $\alpha' \propto 1 / \lambda^{1/2}$. See e.g. \cite{Gubser:1998nz} and numerous subsequent works.} However, one should still think of these corrections as perturbative in powers of $1/\lambda$ due to various potential problems that may arise in theories with higher derivatives, such as the Ostrogradsky instability \cite{Ostrogradsky,Woodard:2015zca}.\footnote{See also \cite{Camanho:2014apa} for a recent discussion of causality violation in theories with higher-derivative bulk actions, in particular with four-derivative, curvature-squared actions.} 

The second source of corrections are the quantum gravity corrections that need to be computed in order to find the $1/N$-corrections in field theory. If we consider $S$ in Eq. \eqref{FullAction} to be a {\em local} quantum effective action, expanded in a gradient expansion, we may also claim that our holographic results incorporate certain types of (perturbative) $1/N$ corrections, included in $\CL$. What is important is the expectation (or the condition) that the anomalous Chern-Simons terms in $\CL_{CS}$ do not renormalise under quantum bulk corrections.

It will prove convenient to write the action \eqref{FullAction} as
\begin{align}
\CL\left[A_a,V_a,G_{ab},\phi_I \right]  \equiv   \CL_G\left[R_{abcd}\right] + \CL_\phi \left[\phi_I\right] +  \CL_A\left[A_a,R_{abcd},\phi_I\right]+
  \CL_V\left[V_a,R_{abcd},\phi_I \right] ,
\label{eqn:general-action}
\end{align}
where $\CL_G$ now contains the Einstein-Hilbert term (along with the cosmological constant) and higher-derivative terms of the
metric, expressed in terms various contractions and derivatives of the Riemann curvature $R_{abcd}$. $\CL_\phi$ contains kinetic and potential terms of a set of neutral scalar fields, $\phi_I$.  By $F_{A,ab}$ and $F_{V,ab}$, we denote the field strengths corresponding to $A_a$ and $V_a$, respectively. Arbitrary derivatives of $F_{A,ab}$ and $F_{V,ab}$ may enter into $\CL_A$ and $\CL_V$, and along with the Chern-Simons terms,
\begin{equation}
\begin{aligned}
\CL_A \left[A_a,R_{abcd},\phi_I\right]&= \CL_A\left[F_{A,ab},\nabla_a F_{A,bc},\ldots,R_{abcd},\nabla_a R_{bcde},\ldots , \phi_I, \partial_a \phi_I, \ldots \right],\\
\CL_V \left[V_a,R_{abcd},\phi_I\right]&= \CL_V\left[F_{V,ab},\nabla_a F_{V,bc},\ldots,R_{abcd},\nabla_a R_{bcde},\ldots , \phi_I, \partial_a \phi_I, \ldots  \right],\\
\CL_{CS} \left[A_a,V_a,G_{ab} \right] &= \epsilon^{abcde} A_a \left( \frac{\kappa}{3} F_{A,bc}F_{A,de} + \gamma
F_{V,bc}F_{V,de} + \lambda R^{p}_{\;\;qbc} R^q_{\;\;pde} \right).
\end{aligned}
\label{eqn:LA-LV-LCS}
\end{equation}
The ellipses `$\ldots$' stand for higher-derivative terms built from $F_{A,ab}$, $F_{V,ab}$,
$R$, $R_{ab}$, $R_{abcd}$ and $\phi_I$.\footnote{Latin letters $\{a,b,c,\ldots\}$ are used
  to label the spacetime indices in the five-dimensional bulk theory while the spacetime
  indices in the dual boundary theory are denoted by the Greek letters
  $\{\mu,\nu,\rho,\ldots \}$. The indices $\{i,j,k,\ldots \}$ represent the spatial
  directions of the boundary theory.} Note also that we have chosen $\CL_A$ and $\CL_V$
so as not to mix the two gauge fields. If there were mixing terms like $F_{A,ab}F_{V}^{ab}$ in the Lagrangian, then the anomalous Ward identities would no longer be those from Eq. \eqref{eqn:non-conserved-current} and additional complications regarding operator mixing would have to be dealt with. We note that the normalisation of the Levi-Civita tensor is chosen to be $\epsilon_{trxyz} = \sqrt{-G}$. 

Our goal is to study coupling constant corrections to the anomalous conductivities that arise from the Ward identity in Eq. \eqref{eqn:non-conserved-current}. We therefore avoid any ingredients in the action \eqref{eqn:general-action} that would explicitly introduce additional terms into \eqref{eqn:non-conserved-current}. Beyond imposing gauge- and diffeomorphism-invariance of \eqref{eqn:non-conserved-current}, we will also restrict our attention to Lagrangians $\CL_A$ and $\CL_V$ that contain no Levi-Civita tensor. An explicit example with violated (bulk) gauge-invariance that can generate a mixed, gauge-global anomaly on the boundary (altering the Ward identity \eqref{eqn:non-conserved-current}) will be studied in Section \ref{section:massive-vector-fields}. 

Furthermore, we assume that the bulk theory admits a homogenous, translationally-invariant and asymptotically anti-de Sitter black brane solution of the form
\begin{equation}
\begin{aligned}
ds^2 &=r^2 f(r) d\bar t^2 + \frac{ dr^2}{r^2g(r)} + r^2 \left(d\bar x^2+d\bar y^2+d\bar z^2\right),\\
A &= A_t(r) d\bar t, \qquad V = V_t(r) d\bar t, \qquad \phi_I = \phi_I (r), 
\end{aligned}
\label{eqn:background-solution-unboosted}
\end{equation}
with $f(r)$ and $g(r)$ two arbitrary functions of the radial coordinate $r$. At AdS infinity,
\begin{align}
\lim_{r\to\infty} f(r) = \lim_{r\to\infty} g(r) = 1.
\end{align}
The coordinates used in Eq. \eqref{eqn:background-solution-unboosted}, $\{\bar x^\mu,r\}$, will be referred to as the un-boosted coordinates. Near the (outer) horizon, we assume that the metric can be written in a non-extremal, Rindler form  
\begin{align}
f(r) &= f_1 (r-r_h) + f_2 (r-r_h)^2+\CO(r-r_h)^3, \label{nearH1}\\
g(r) &= g_1(r-r_h) + g_2 (r-r_h)^2+ \CO(r-r_h)^3. \label{nearH2}
\end{align}
The Hawking temperature of this black brane background (and its dual) is given by
\begin{equation}
T = \frac{r_h^2}{4\pi} \sqrt{ f_1 g_1}.
\label{eqn:temperature}
\end{equation}

The classical equations of motion describing this system can be obtained by varying the
action \eqref{eqn:general-action}. Firstly, the variations of the two gauge fields
give\footnote{In five spacetime dimensions, we define the Hodge dual of a $p$-form $\Omega =
  (p!)^{-1}\Omega_{a_1\ldots a_p}dx^{a_1}\wedge \ldots\wedge dx^{a_p}$ as 
\begin{equation}\nonumber
\star \Omega = \frac{1}{p!(5-p)!} \sqrt{-G}\;
\Omega_{a_1\ldots a_p}\epsilon^{a_1\ldots a_p}_{\;\;\;\;\;\;\;\;\;\;\; a_{p+1} \ldots a_{5}}
dx^{a_{p+1}}\wedge\ldots \wedge dx^{a_5}.
\end{equation} }
\begin{align}
d\star H_5 = 0, && d\star H = 0 ,
\label{eqn:Maxwell-eoms}
\end{align}
where the two-forms $H_5$ and $H$ are defined as 
\begin{equation}
\begin{aligned}
H_5 &= \frac{1}{2}\left( \frac{\delta \left( \CL_A\right)}{\delta \left(\grad^a A^b\right)}- \grad_c \frac{\delta \left( \CL_A\right)}{\delta
\left(    \grad_c \grad^a A^b\right)} + \ldots \right) dx^a dx^b + \kappa \star \omega_A + \gamma\star
\omega_V + \lambda \star \omega_\Gamma , \\
H &= \frac{1}{2}\left( \frac{\delta \left( \CL_V\right)}{\delta \left(\grad^a V^b\right)}- \grad_c \frac{\delta \left( \CL_V\right)}{\delta
 \left(   \grad_c \grad^a V^b\right)} + \ldots \right) dx^a dx^b + \gamma \star(V\wedge dA) . 
\end{aligned}
\label{eqn:2-form-H}
\end{equation}
The ellipses again denote expressions coming from the higher-derivative terms. The three abelian Chern-Simons three-forms are composed of the two gauge field one-forms $A = A_a dx^a$ and
$V= V_a dx^a$, and the Levi-Civita connection one-form $\Gamma^a_{~b} = \Gamma^a_{~bc}\,
dx^c$ as  
\begin{align}
\omega_X = \Tr \left( X \wedge dX + \frac{2}{3} X\wedge X \wedge X  \right), 
\label{eqn:Chern-Simons-3-form}
\end{align}
where $X = \{ A ,V , \Gamma^a_{~b}  \}$.\footnote{In terms of the index notation, the Chern-Simons form built out of the Levi-Civita connection is given by  
\begin{equation}\nonumber 
\omega_{abc} = \Gamma^{p_1}_{\;\;p_2a }\partial_b \Gamma^{p_2}_{\;\;p_1
      c} + (2/3)\Gamma^{p_1}_{\;\;p_2 a} \Gamma^{p_2}_{\;\;p_3b} \Gamma^{p_3}_{\;\;p_1 c}. 
\end{equation}}

Secondly, varying the metric gives the Einstein's equation 
\begin{equation}
R_{ab} - \frac{1}{2} G_{ab} R + \ldots  = T^{M}_{ab} + \frac{1}{2}\grad_c
\left(\Sigma_{ab}^{~~c}+ \Sigma_{ba}^{~~c} \right),
\label{eqn:Einstein-eoms}
\end{equation}
where $T^M_{ab}$ is the stress-energy tensor for the scalars and the gauge fields, excluding the Chern-Simons terms. The {\em spin current} $\Sigma_{ab}^{~~c}$ is defined as 
\begin{equation}
\Sigma_{ab}^{~~c} = -\lambda\; \epsilon_a^{\,\,\, d_1d_2d_3d_4}F_{d_1d_2} R_{d_3d_4
  b}^{~~~~~~ c}.
\end{equation}
We refer the reader to \cite{Azeyanagi:2013xea} for a more general definition of the
spin current, its connection to the anomaly polynomial in Eq. \eqref{eqn:anomaly-polynomial} and expressions for $\Sigma_{ab}^{~~c}$ for different anomaly polynomials. We assume that the equations of motion coming from the variations of the scalar fields in \eqref{FullAction} can also be solved, but we will make no further reference to that set of equations. As stated above, the full system of equations is assumed to result in a non-extremal, asymptotically AdS black brane solution and non-trivial, backreacted profiles for the gauge and the scalar fields.  

To find the set of anomalous conductivities $\{ \sigma_{J_5 B}, \sigma_{J B}, \sigma_{J_5 \omega},
\sigma_{J \omega}\}$ in all hypothetical duals of this holographic setup, it is convenient to consider the following
perturbed metric in the boosted (fluid-gravity) frame \cite{Azeyanagi:2013xea}:
\begin{equation}
ds^2 = -2 \sqrt{\frac{f(r)}{g(r)}} u_\mu dr dx^\mu + r^2 f(r) u_\mu u_\nu dx^\mu dx^\nu +r^2
\Delta_{\mu\nu} dx^\mu dx^\nu + 2 r^2 h(r) u_\mu \omega_\nu dx^\mu dx^\nu ,
\label{eqn:perturbed-metric}
\end{equation}
where the projector $\Delta_{\mu\nu}$ is defined as $\Delta_{\mu\nu} = \eta_{\mu\nu} +
u_\mu u_\nu$, with $\eta_{\mu\nu}$ the four-dimensional Minkowski metric.  Note that once we set the
fluid to be stationary, i.e. $u^\mu_{eq} = \{-1,0,0,0\}$, the metric
\eqref{eqn:perturbed-metric} will return to the un-boosted form
\eqref{eqn:background-solution-unboosted}, but in the Eddington-Finkelstein coordinates,
as is usual in the fluid-gravity correspondence
\cite{Bhattacharyya:2008jc,Rangamani:2009xk}. The perturbations are organised so that the fluid velocity $u_\mu$ depends only on the boundary coordinates $x^\mu$ and all of the $r$-dependence is encoded in $h(r)$. Since the vorticity is defined as $\omega^\mu =
\epsilon^{\mu\nu\rho\sigma} u_\nu \d_\rho u_\sigma$, the last term in
\eqref{eqn:perturbed-metric} corresponds to the metric perturbations at first
order in the derivative expansion (in the $x^\mu$ coordinates). Similarly, the perturbed
axial and vector gauge fields can be written as\footnote{Our choice of the metric and the gauge fields can be understood in the following way: If one considers the perturbed metric and the gauge fields with all possible terms at first order in gradient expansions, they have the form
\begin{align}\nonumber
ds^2 =-2 S(r) \,u_\mu dx^\mu dr+ F(r)\, u_\mu u_\nu dx^\mu dx^\nu + G(r)\,
\Delta_{\mu\nu}dx^\mu dx^\nu + 2 H^\perp_\mu(r,x) \,u_\nu dx^\mu dx^\nu+
\Pi(r)\,\sigma_{\mu\nu} dx^\mu dx^\nu,
\end{align}
\begin{align}\nonumber
&A = C(r) u_\mu dx^\mu + a^\perp_\mu(r,x) dx^\mu ,&& V = D(r) u_\mu dx^\mu + v^\perp_\mu(r,x) dx^\mu,
\end{align}
where $H^\perp_\mu$, $a^\perp_\mu$ and $v^\perp_\mu$ are vectors orthogonal to the fluid
velocity $u^\mu$. Using the equations of motion for $\{ H^\perp_\mu,
a^\perp_\mu, v^\perp_\mu\}$, one can show that they decouple from all other perturbations at the same
order in the gradient expansion (see e.g. \cite{Erdmenger:2008rm,Banerjee:2008th}). Thus, to compute anomalous conductivities, one can consistently solve for only $\{ H^\perp_\mu,
a^\perp_\mu, v^\perp_\mu\}$, setting the remaining perturbations to zero. To first order, this gives our Eqs. \eqref{eqn:perturbed-metric} and \eqref{eqn:perturbed-gauge-fields}. 
}
\begin{equation} 
\begin{aligned}
&A = -A_t (r)\, u_\mu dx^\mu + \tilde{a}(x^\mu) + a(r) \,\omega_\mu dx^\mu,\\
&V = -V_t (r)\, u_\mu
dx^\mu + \tilde{v}(x^\mu) + v(r) \,\omega_\mu dx^\mu.
\label{eqn:perturbed-gauge-fields}
\end{aligned}
\end{equation}
One may use the one-forms $\tilde{a}$ and $\tilde{v}$ to define the magnetic field source $B^\mu =
\epsilon^{\mu\nu\rho\sigma} u_\nu \d_\rho \tilde{v}_\sigma$ and the (fictitious) axial magnetic field source $B_5^\mu =
\epsilon^{\mu\nu\rho\sigma} u_\nu \d_\rho \tilde{a}_\sigma $.

\section{Proof of universality}
\label{section:anomaly-induced-electric-current}

In this section, we show that upon expanding the equations of motion \eqref{eqn:Maxwell-eoms} and \eqref{eqn:Einstein-eoms} to first order in the (boundary) derivative expansion, the conserved currents can be expressed as a total radial derivative of some function. This type of a radially conserved quantity is necessary for the applicability of the membrane paradigm, used e.g. in \cite{Iqbal:2008by} and many other holographic studies. To express all four anomalous conductivities purely in terms of the near-horizon data, our work will generalise the membrane paradigm result for the chiral magnetic conductivity of G\"{u}rsoy and Tarr\' io \cite{Gursoy:2014boa}. This will then enable us to establish the universality of the four transport coefficients in the presence of a general higher-derivative bulk theory specified in Section \ref{section:holographic setup}. Furthermore, the structure of the equations will single out the properties that holographic theories must violate in order for there to be a possibility that the dual conductivities may get renormalised. 

Our proof can be divided into two steps: First (in Section \ref{sec.MemParad}), we expand the equations of motion for the gauge field \eqref{eqn:Maxwell-eoms} to first order in the (boundary coordinate) derivative expansions and arrange them into a total-derivative form of a conserved current along the radial direction. This radially conserved current can be written as a sum of the anomalous Chern-Simons terms and terms that come from the rest of the action. We identify the conditions that each of these terms has to satisfy in order for the anomalous conductivities to have a universal form fixed by the Chern-Simons action. Proving the validity of these conditions is then done in Section \ref{section:any-higher-derivative} by analysing the horizon and the boundary behaviour of the higher-derivative bulk effective action (and all possible resulting terms that can appear in the conserved current).

\subsection{Anomalous conductivities and the membrane paradigm}
\label{sec.MemParad}

Let us begin by considering the axial and the vector currents, $\la \delta J^\mu_5 \ra$ and $\la
\delta J^\mu \ra$, sourced by a small magnetic field and a small vortex. As in \cite{Gursoy:2014boa}, the membrane paradigm equations follow from the two Maxwell's equations in \eqref{eqn:Maxwell-eoms}. For conciseness, we only show the details of the axial current computation, which involves $H_5$ from Eq. \eqref{eqn:2-form-H}. A calculation for the vector current, involving $H$, proceeds along similar lines. In case of the vector current, we will only state the relevant results. 

To first order in the gradient expansion along the boundary directions $x^\mu$, both equations in \eqref{eqn:Maxwell-eoms} can be schematically written as
\begin{equation}
\d_r \left(\sqrt{-G} H^{ra}_5\left(\d^1\right) \right)+ \d_\mu \left( \sqrt{-G} H^{\mu a}_5\left(\d^0\right)\right) = 0, 
\label{eqn:expand-Maxwell}
\end{equation}
where $H^{ra}_5\left(\d^0\right)$ and $H^{\mu a}_5\left(\d^1\right)$ are the components of the conserved current two-form in Eq. \eqref{eqn:2-form-H} that contain zero- and one-derivative terms (derivatives are taken with respect to $x^\mu$).

As our first goal is to rewrite the problem in terms of a radially conserved quantity, we need to consider the structure of second term in \eqref{eqn:expand-Maxwell}. We will set the index $a$ to the four-dimensional index $\nu$. It is easy to see that only the Chern-Simons terms from $\CL_{CS}$ can enter into this term at zeroth order in the (boundary) derivative expansion, i.e. $ \d_\mu \left(\sqrt{-G} H_5^{\mu\nu}\left(\d^0\right)\right)\vert_{\kappa=g=\lambda=0}=0$ (cf. Eq. \eqref{eqn:LA-LV-LCS}). This is because $H_5^{\mu\nu}$ can only be constructed out of the (axial) gauge field \eqref{eqn:perturbed-gauge-fields} and the metric tensor \eqref{eqn:perturbed-metric}, containing no derivatives along $x^\mu$. At zeroth-order in the derivative expansion, any two-tensor $X^{\mu\nu}$ can thus be decomposed as
\begin{equation}
X^{\mu\nu} = X_1 \, u^\mu u^\nu + X_2 \, \Delta^{\mu\nu} + X_3 \, u^{(\mu} A^{\nu)} + X_4\, u^{[\mu} A^{\nu]} ,
\end{equation}
where $X_i$ are scalar functions of the radial coordinate. For an anti-symmetric $X^{\mu\nu}$, as are $H_5^{\mu\nu}$ and $H^{\mu\nu}$, $X_1$, $X_2$ and $X_3$ must vanish and only $X_4$ can be non-zero. Since such a term can only come from $\CL_{CS}$, $\CL$ cannot contribute to the second term in \eqref{eqn:expand-Maxwell}. For $a = \nu$, the two terms in Eq. \eqref{eqn:expand-Maxwell} are therefore given by 
\begin{equation}
\begin{aligned}
\d_r \left[ \sqrt{-G} H^{r\nu}_5\left(\d^1\right) \right] & = \frac{\partial}{\partial r}\left[\ldots + \kappa \left(A_t B_5^\nu + A_t^2 \omega^\nu \right) + \gamma \left(V_t B^\nu + V_t^2 \omega^\nu \right) + \lambda \frac{g(r^3f' )^2}{2r^2 f} \omega^\nu \right] ,\\
\d_\mu \left[\sqrt{-G} H^{\mu \nu}_5 \left(\partial^0\right) \right] &= \kappa\left(\partial_r A_t\right)  B_5^\nu  + \gamma \left( \partial_r V_t \right) B^\nu  = \partial_r \left( \kappa A_t B^\nu_5 + g V_t B^\nu \right).
\end{aligned}
\label{eqn:CS-part-of-H}
\end{equation}
The ellipsis indicates the non-Chern-Simons terms. Hence, one can write the Maxwell's equation for the axial gauge field as a derivative of a conserved current along the $r$-direction:
\begin{equation}
\partial_r \CJ^\mu_5 (r) = 0.
\label{eqn:membrane-current-J5}
\end{equation}
The axial bulk current is defined as
\begin{equation}
\CJ^\mu_5 (r)= \CJ^{\mu}_{5,mb}(r) + \CJ^{\mu}_{5,r}(r) +
\CJ^\mu_{5,CS}(r) ,
\label{eqn:def-CJ5}
\end{equation}
where the {\em membrane current} $\CJ^\mu_{5,mb}(r)$, the Chern-Simons current
$\CJ^\mu_{5,CS}$ and $\CJ^\mu_{5,r}$ are defined as
\begin{equation}
\begin{aligned}
\CJ^\mu_{5,mb} &= \sqrt{-G} \left( \frac{\d \CL_A}{\d A'_\mu}
- \d_a \frac{ \d  \CL_A }{\d(\d_a A'_\mu)} + \ldots \right)\biggr\vert_{h(r) \to 0} , \\
\CJ^\mu_{5,r} &=  \sqrt{-G} \left( \frac{\d \CL_A }{\d A'_\mu}
- \d_a \frac{ \d \CL_A}{\d(\d_a A'_\mu)}+\ldots \right)\biggr\vert_{a(r)\to 0} , \\
\CJ^\mu_{5,CS} &= 2\kappa A_t B^\mu_5 + 2\gamma V_t B^\mu + \left( \kappa A_t^2 + \lambda
  \frac{g(r^2 f')^2}{2f } \right)\omega^\mu\, .
\end{aligned}
\label{eqn:membrane-and-CS-current}
\end{equation}
Note that the primes indicate derivatives with respect to the radial coordinate. 

The expectation value of the external boundary current $\la \delta J^\mu_5\ra$ that we turned on to excite anomalous transport (cf. Eq. \eqref{CondDef}) is obtained by varying the perturbed on-shell action \eqref{eqn:general-action} with respect to the bulk axial gauge field fluctuation at the boundary. We find that it is the membrane current $\CJ^\mu_{5,mb}$ evaluated at the boundary ($r\to \infty$) that can be interpreted as its expectation value:
\begin{equation}\label{ExpJasJmb}
\la \delta J^\mu_5 \ra =  \lim_{r\to \infty} \CJ^\mu_{5,mb}(r).
\end{equation}
This result is of central importance to the existence of the membrane paradigm in our discussion. 

Let us now study how $\CJ^\mu_{5,mb}$ can be related to the full conserved current $\CJ^\mu$ from Eq. \eqref{eqn:def-CJ5}. What will prove very convenient is the gauge choice for $A$ and $V$ whereby (see e.g. \cite{Gynther:2010ed})
\begin{align}
\lim_{r\to\infty} A_t(r )=0, && \lim_{r\to\infty} V_t(r)=0 .
\end{align} 
Such a choice results in\footnote{For an alternative gauge
choice, see e.g. formalism B from Ref. \cite{Landsteiner:2012kd}.} 
\begin{align}
\lim_{r\to \infty} \CJ_{5,CS} (r) = 0,
\end{align}
which together with the conservation equation \eqref{eqn:membrane-current-J5} and Eq. \eqref{ExpJasJmb} implies that
\begin{align}\label{eqn:vev-J5}
\la \delta J^\mu_5 \ra = \CJ^{\mu}_{5,mb}(r_h) + \CJ^{\mu}_{5,r}(r_h) -  \CJ^\mu_{5,r}(\infty) +
\CJ^\mu_{5,CS}(r_h).
\end{align}
What we will prove in the next section (Sec. \ref{section:any-higher-derivative}) will be the statement that for any theory specified by the action in \eqref{FullAction},
\begin{align}\label{eqn:condition-for-universality}
\CJ^{\mu}_{5,mb}(r_h) + \CJ^{\mu}_{5,r}(r_h) -  \CJ^\mu_{5,r}(\infty) = 0,
\end{align}
implying that the current $\la \delta J^\mu_5\ra$ can be completely determined by only the
Chern-Simons current evaluated at the horizon, 
\begin{align}\label{MembraneParadigmFullCurrentFinal}
\la \delta J^\mu_5 \ra &=  \CJ^\mu_{5,CS}(r_h).
\end{align}
The same reasoning and equations \eqref{ExpJasJmb}--\eqref{MembraneParadigmFullCurrentFinal} apply also to the case of the vector current, up to the appropriate replacements of $A_a$ by $V_a$, $\CL_A$ by $\CL_V$ and the axial Chern-Simons current by
\begin{equation}
\CJ^\mu_{CS} = 2\gamma \left( A_t B^\mu + V_t B^\mu_5\right)+ 2\gamma A_t V_t \, \omega^\mu.
\end{equation}

Let us for now assume that the condition \eqref{eqn:condition-for-universality} is satisfied and proceed to compute the anomalous conductivities. In our gauge choice, the gauge fields at the horizon are related to the two chemical potentials via 
\begin{align}\label{AVhorizon}
A_t(r_h) = -\mu_5, && V_t(r_h) = -\mu.
\end{align}
By using the near-horizon expansions \eqref{nearH1} and \eqref{nearH2}, the last term in $\CJ^\mu_{5,CS}$ from \eqref{eqn:membrane-and-CS-current} can be related to the temperature 
\begin{equation}
\frac{g \left(r^2f'\right)^2}{f} = r^4 f_1 g_1 =4 \left(2\pi T\right)^2.
\end{equation}
Furthermore, using the horizon values of the gauge fields from Eq. \eqref{AVhorizon} along with the definitions of the anomalous conductivities from \eqref{CondDef}, we find
\begin{align}
&\sigma_{J_5 B} = -2 \gamma \mu, && \sigma_{J B} = -2 \gamma \mu_5,\nn
& \sigma_{J_5 \omega} = \kappa \mu_5^2 +
\gamma \mu^2 + 2\lambda (2\pi T)^2, && \sigma_{J \omega} = 2\gamma \mu_5 \mu .
\label{eqn:universal-anomalous-conductivity-electric}
\end{align}
Hence, so long as the condition \eqref{eqn:condition-for-universality} is satisfied, the bulk theory \eqref{FullAction} gives precisely the non-renormalised, universal conductivities stated in Eqs. \eqref{s1}, \eqref{s2} and \eqref{s3}.

\subsection{Universality}
\label{section:any-higher-derivative}

We will now show that the condition \eqref{eqn:condition-for-universality} always holds in theories in which $\CL$ (as defined in Eq. \eqref{FullAction}) is gauge- and diffeomorphism-invariant. Thus, we will establish the universality of the anomaly-induced conductivities $\sigma_{J_5 B}$, $\sigma_{J B}$, $\sigma_{J_5 \omega}$ and $\sigma_{J \omega}$ from Eq. \eqref{eqn:universal-anomalous-conductivity-electric} in theories with arbitrary higher-derivative actions, dual to an infinite series of coupling constant corrections expanded around infinite coupling. The condition \eqref{eqn:condition-for-universality} requires us to understand how $\CJ_{5,mb}^\mu$ and $\CJ_{5,r}^\mu$ behave at the two ends of the five-dimensional geometry (boundary and horizon). To make general statements about that, we construct an effective field theory (or the effective current) in terms of the metric, gauge fields and dilatons with first-order perturbations to quadratic order in the amplitude expansion. The two conditions that we impose on the effective theory and the resulting currents are the following: 
\begin{itemize}
\item[(1)] The theory must be regular at the non-extremal horizon, by which we mean that any Lorentz scalar present in the action (or a current) must be regular (non-singular) when evaluated at the horizon.
\item[(2)] The bulk spacetime is asymptotically anti-de Sitter.
\end{itemize}
For conciseness, we again only analyse the axial gauge field, $A_a$. A completely equivalent
procedure can be applied to the case of the vector gauge field, $V_a$.   

From the definitions of $\CJ^\mu_{5,mb}$ and $\CJ^\mu_{5,r}$ in Eq. \eqref{eqn:membrane-and-CS-current}, it is clear that the only
relevant part of the action \eqref{eqn:general-action} for this analysis is $\CL_A$. Because the two currents are independent of the Chern-Simons terms, they only depend on the terms
encoded in $H^{ra}_{5}\left(\d^1\right)$ (see discussion below Eq. \eqref{eqn:expand-Maxwell}). The possible terms in $H^{ra}_5\left(\d^1\right)$ that correspond to $\CJ^\mu_{5,mb}$ and $\CJ^\mu_{5,r}$ can be written (schematically, up to correct tensor structures of $ \CC_{A,n} $ and $ \CC_{G,n}$) as
\begin{equation}
H^{r\mu}_{5}\left(\d^1\right) = \sum_{n=1}^\infty \left[ \CC_{A,n} \d^n_r a(r) + 
  \CC_{G,n}\d^n_r h(r) \right] \omega^\mu + H^{r\mu}_{5,CS}\left(\d^1\right),
\label{eqn:non-CS-H}
\end{equation}
where $H^{r\mu}_{5,CS}$ is the irrelevant Chern-Simons part of $H^{r\mu}_5$, stated explicitly in Eq. \eqref{eqn:CS-part-of-H}. Since the action $\CL_A$ does not contain any Levi-Civita tensors, the terms in $\{\CC_{A,n},\,\CC_{G,n}\}$ can only depend on $a(r)$ and $h(r)$. This implies that $\CC_{A,n} = \CC_{G,n} = 0$ when $a(r)=h(r)=0$, to first order in the boundary-coordinate derivative expansion. Hence, the problem reduces to the question of finding all possible structure of the tensorial coefficients $\{\CC_{A,n},\,\CC_{G,n}\}$ at the horizon and at the boundary. 

It is now convenient to return to the un-boosted coordinates, $\{ r, \bar{x}^\mu\}$, used in Eq. \eqref{eqn:background-solution-unboosted}. In these coordinates, the perturbed metric and the axial gauge field are (in analogy with
\eqref{eqn:perturbed-metric} and \eqref{eqn:perturbed-gauge-fields})
\begin{align}
ds^2 &= -r^2f(r) d\bar{t}^2 + \frac{d r^2}{r^2g(r)} + r^2 (d\bar x^2 + d\bar y^2 + d \bar
z^2) + 2h_{\bar t  i}(r,\bar x^i) d\bar t d\bar x^i,\\
A &= A_t dt + a_i(r,\bar x^i) d\bar x^i,
\end{align}
where the perturbations are now denoted by $h_{\bar t  i},$ $a_i$ and $v_i$ with
$i=\{x,y,z\}$. One can relate $\{ h_{\bar t  i}, a_i\}$ to $\{h(r), a(r)\}$ by using the appropriate coordinate transformations, which give
\begin{align}
h_{\bar t i} &= \ldots +r^2 h(r) \, u_\mu \omega_\nu \frac{\d x^\mu}{\d \bar{t}} \frac{\d x^\nu}{\d
  \bar{x}^i} + \CO\left(\d^2\right), \nn
a_{i} &= \ldots + a(r) \, \omega_\mu \frac{\d x^\mu}{\d \bar
  x^i} + \CO\left(\d^2\right).
\label{eqn:boosted-h-and-a}
\end{align}
Here, the ellipses denote the zeroth-order terms in the derivative expansion. It is convenient to consider $u^\mu-u^\mu_{eq}$ to be small, which gives
\begin{align}
u_\mu dx^\mu = dt + \delta u_i dx^i,&& dt = d\bar t +
\frac{1}{r^2}\sqrt{\frac{1}{f(r)g(r)}} dr, && dx^i = d\bar x^i.
\label{eqn:coordinate-transformation}
\end{align}
This choice of the fluid velocity further gives $\omega^t = B^t = 0$. Thus, in the remainder in this section, we will only write
down the tensors $\{H^{r\mu}_5,\CJ^\mu_5,\CJ^\mu_{5,CS} \}$ with spatial components of $\mu = \{i,j,k,\ldots \}$. It immediately follows that $H^{ri}_5 \left(r,x^\mu\right)$ in the boosted coordinates and $H^{ri}_5\left(r,\bar x^\mu\right)$ in the un-boosted coordinates have identical expressions. In analogy with \eqref{eqn:non-CS-H}, expanding $H_5^{ri}$ in the un-boosted coordinates to first order in amplitudes of $a_i$ and $h_{\bar t i}$,
\begin{equation}
\begin{aligned}
H_5^{ri}\left[a_i,h_{ti}\right] &= \left( \CI^{rirj}_{A,1}\d_r a_j +\CI^{rirrj}_{A,2} \d^2_r a_j
  + \ldots \right) + \left( \CI_{G,0}^{ri\bar t j} h_{\bar t j} + \CI_{G,1}^{r i \bar t r
    j}\d_r h_{\bar t j} + \CI_{G,2}^{ri\bar t r r j}\d^2_r h_{\bar t j}+ \ldots \right)\\
&\quad + \left( \text{terms with derivatives along $x^i$}\right).
\end{aligned}
\label{eqn:delta-H-unboosted}
\end{equation}
Note that $\CI^{rij}_{A,0}=0$ because gauge-invariance of $\CL_A$ excludes the possibility of any explicit dependence on $a_i$ (only derivatives of $a_i$ may appear). The ellipses represent terms with higher derivatives in $r$ and $\{\CI_{A,n},\CI_{G,n}\}$ are tensors contracted with $\d^n_r a_i$ and
$\d^n_r h_{\bar t i}$. To verify \eqref{eqn:delta-H-unboosted}, we can use the coordinate transformations \eqref{eqn:boosted-h-and-a}, which show that all relevant terms from \eqref{eqn:non-CS-H} are indeed contained in \eqref{eqn:delta-H-unboosted}. Thus, one can determine the coefficients $\{\CC_{A,n},\CC_{G,n}\}$ by applying \eqref{eqn:coordinate-transformation} to \eqref{eqn:delta-H-unboosted} and matching the coefficients of $\d^n_ra(r)\,\omega^i$ and $\d^n_r h(r) \,\omega^i$.

The structure of the $\{\CI_{G,n}, \CI_{A,n}\}$ tensors near the horizon and the AdS-boundary can be understood in the following way: In the un-boosted frame, we define five mutually orthogonal unit-vectors or vielbeins, $e_{\hat{p}a} = \delta_{\hat p a}$, where the hatted indices $\{ \hat{p},\hat
q,..\} = \{\hat 0,\hat1,\hat2,\hat3,\hat4\}$ are used as (local flat space) bookkeeping indices. The full set of the five-dimensional vectors with upper Lorentz indices can now be written as $e^a_{\;\;\hat p} = \left[\sqrt{G}\right]^{ab} \delta_{\hat p b}$:
\begin{equation}
\begin{aligned}
e_{\hat 0} &= \Big(\left(r^2f\right)^{-1/2},0,0,0,0\Big),\\
e_{\hat 1} &= \Big(0,1/r,0,0,0\Big),\\
e_{\hat 2} &= \Big(0,0,1/r,0,0\Big),\\
e_{\hat 3} &= \Big(0,0,0,1/r,0\Big),\\
e_{\hat 4} &= \Big(0,0,0,0,\left(r^2g\right)^{1/2}\Big).
\end{aligned}
\label{eqn:vielbein-components}
\end{equation}
These normal vectors allow us to write the tensors $\{ \CI_{G,n},\CI_{A,n}\}$ as  
\begin{equation}
\begin{aligned}
\CI^{a_1a_2\ldots a_m}_{A,n} &= \sum_{\hat{p}_1,\ldots ,\hat{p}_m} \CS_{A,n}^{\hat p_1\ldots \hat p_m}
e^{a_1}_{\;\;\hat p_1} \ldots e^{a_m}_{\;\;\hat p_m}, \\
\CI^{a_1a_2\ldots a_m}_{G,n} &= \sum_{\hat{p}_1,\ldots ,\hat{p}_m} \CS_{G,n}^{\hat p_1\ldots \hat p_m}
e^{a_1}_{\;\;\hat p_1} \ldots e^{a_m}_{\;\;\hat p_m},
\end{aligned}
\label{eqn:decomposed-CC}
\end{equation}
where $\{\CS_{A,n},\CS_{G,n}\}$ are (spacetime) Lorentz-scalars. The regularity condition imposed at the horizon demands that these scalar have to be non-singular at $r=r_h$. The question of whether $\CI_{G,n}$ and $\CI_{A,n}$ vanish at the horizon is therefore completely determined by the values the projectors $e^{a_1}_{\;\;\hat p_1} \ldots e^{a_m}_{\;\;\hat p_m}$ take when evaluated at the horizon. To demonstrate this fact more clearly, let us write down the first few relevant components of the tensors $\CI_{G,n}$ and $\CI_{A,n}$ explicitly:
\begin{align*}
&\CI_{G,0}^{ri \bar t j}=\left(r^{-2}\sqrt{g/f}\right)\,\CS_{G,0}^{4\hat i 0 \hat j}\;,
 & &\CI_{A,0}^{rij} = 0\;,\\
&\CI_{G,1}^{ri \bar t rj}=\left(r^{-1}\sqrt{g^2/f}\right)\,\CS_{G,2}^{4\hat i 04\hat
  j}\;, &&\CI_{A,1}^{ri rj} = g \, \CS^{4\hat i 4\hat j}_{A,1}\;,\\
&\CI_{G,2}^{ri\bar t rrj}=\left(\sqrt{g^3/f}\right) \,\CS_{G,2}^{4\hat i 044\hat
  j}\;, &&\CI_{A,2}^{rirrj}= \left(r g^{3/2}\right) \CS_{A,2}^{4\hat i 44\hat j}\;,\\
&\CI_{G,3}^{ri\bar t rrrj}=\left(r\sqrt{g^4/f}\right)\,\CS_{G,3}^{4\hat i 0444\hat j}\;,&&\CI_{A,3}^{rirrrj}= \left( r^2g^2 \right)\, \CS_{A,3}^{4\hat i 444\hat j}\;,
\end{align*}
with $r = r_h$. As before, the tensor $\CI^{rij}_{A,0}=0$ because of the gauge-invariance of $\CL_A$.

With this decomposition, the problem of determining the non-zero terms in $H^{ri}_5$ has been reduced to simple power-counting. Namely, a tensor $\CI^{a_1a_2\ldots}$ can only be non-zero at the horizon if the number of $e^{\bar{t}}_{\;\hat 0}$ in its decomposition is equal to or greater than the number of $e^{r}_{\;\hat 4}$. The regularity of the scalars $\CS_{A,n}$ and $\CS_{G,n}$ at the horizon plays a crucial role here. Hence, one can see that the only non-zero tensor from the set of $\{\CI_{A,n},\CI_{G,n}\}$ is $\CI_{G,0}^{ri\bar t j}$. The conserved current evaluated at the horizon thus becomes
\begin{equation}
\CJ^i_5  = \sqrt{-G}\left(\sqrt{\frac{g}{f}} \;\CS^{4\hat j 0 \hat
  i}_{G,0}\right) h(r_h) \,u_\mu \omega_\nu \frac{\d x^\mu}{\d \bar t} \frac{\d x^\nu}{\d \bar x^j} + \CJ^i_{5,CS} (r_h) .
\label{eqn:non-CS-CJ5-general}
\end{equation}
To see why the first term in \eqref{eqn:non-CS-CJ5-general} has to vanish, recall that as other scalars, the Ricci scalar also has to be regular at the horizon. As pointed out in \cite{Gursoy:2014boa}, this condition implies that $h_{ti} \sim (r-r_h)$ at the horizon. Therefore, the conserved current at the horizon is indeed fully determined by the anomalous Chern-Simons term:
\begin{align}\label{CJ5onlyCSmatters}
\CJ^i_5 =  \CJ^i_{5,CS} (r_h).
\end{align}
With $H^{rt}_5 = 0$, Eq. \eqref{CJ5onlyCSmatters} implies the first two terms from the condition \eqref{eqn:condition-for-universality} vanish: 
\begin{align}\label{ProofIng1}
\CJ^{\mu}_{5,mb}(r_h) + \CJ^{\mu}_{5,r}(r_h)  = 0.
\end{align}

Similarly, we can determine the value of the current $\CJ^\mu_{5,r}$ at the boundary. Since $\CJ^\mu_{5,r}$ includes only terms linear in $h(r)$, it is enough to consider 
\begin{equation}
H^{ri}_5  = \left( \CI_{G,0}^{ri\bar t j} h_{\bar t j} + \CI_{G,1}^{r i \bar t r
    j}\d_r h_{\bar t j} + \CI_{G,2}^{ri\bar t r r j}\d^2_r h_{\bar t j}+ \ldots \right) + \ldots.
\label{eqn:delta-H-unboosted-Jr}
\end{equation}
Now, because the boundary is asymptotically AdS and higher-derivative terms considered here
do not change the scaling behaviour near the boundary, we can use the near-AdS solution for $h(r)$ \cite{Azeyanagi:2013xea}:
\begin{align}
h(r) = \frac{\CH}{r^4} + \CO\left(r^{-5}\right) .
\label{eqn:near-boundary-h}
\end{align}
Substituting the expansion for $h(r)$ into \eqref{eqn:delta-H-unboosted-Jr}, it immediately follows that the third term in the condition \eqref{eqn:condition-for-universality} vanishes as well when it is evaluated at the boundary (note again that $H^{rt}_5 = 0$):
\begin{align}\label{ProofIng2}
\CJ^\mu_{5,r} (\infty) = 0.
\end{align}

Together, Eqs. \eqref{ProofIng1} and \eqref{ProofIng2} imply the validity of the condition stated in Eq. \eqref{eqn:condition-for-universality}, which completes our proof. The analysis of the vector current $\CJ^\mu$ and a proof of a condition analogous to \eqref{eqn:condition-for-universality} follow through along exactly the same lines. This implies that all four anomalous conductivities take the universal form of \eqref{eqn:universal-anomalous-conductivity-electric} for all holographic theories specified in \eqref{FullAction} so long as the (effective) theory is regular at the non-extremal horizon and the bulk is asymptotically anti-de Sitter.

\section{Examples and counter-examples}
\label{section:examples}

In this section, we turn our attention to explicit examples of theories that obey and violate the conditions used in our proof in Section \ref{section:anomaly-induced-electric-current} and thus result in universal and renormalised anomalous conductivities, respectively. We will first demonstrate their universality in two- and four-derivative theories with a non-extremal horizon and then move on to describing two holographic models, which violate the assumptions in the proof of Eq. \eqref{eqn:condition-for-universality}. More precisely, in Section
\ref{section:Einstein-Hilbert-Maxwell-gravity}, we compute the conductivities in the two-derivative Einstein-Maxwell-Dilaton theory. In Section \ref{section:4-derivatives}, we then show explicitly how our proof works in the case of the most general four-derivative action with Maxwell fields and dynamical gravity. In both of those case, the conductivities are universal and the current at the horizon only depends on the metric fluctuation, as established by our
effective theory method in \eqref{eqn:non-CS-CJ5-general}. 

In Section \ref{section:counter-examples}, we comment on the validity of our proof in gravity duals without a horizon. We use the examples of the confining soft/hard-wall models and charged dilatonic black holes at zero temperature. The membrane paradigm computation goes through as before in the case of confining geometry. However, the conductivities no longer have any temperature dependence, which would require us to augment the replacement rule discussed in Appendix \ref{app:anomP}. As for the latter example, the family of theories considered suffers from naked singularities in the bulk. Lastly, in Section \ref{section:massive-vector-fields}, we point out how the bulk terms corresponding to field theories with a gauge-global anomaly violate the assumptions in our proof. This is consistent with the known fact that anomalous conductivities in systems with mixed anomalies receive corrections along the renormalisation group flow. We will not review the details behind the holographic constructions of such systems but rather focus on the reasons for why these models may violate the universality from the point of view of Section \ref{section:any-higher-derivative}.      

\subsection{Einstein-Maxwell-dilaton theory at finite temperature}
\label{section:Einstein-Hilbert-Maxwell-gravity}

As for our first example, we consider the two-derivative Einstein-Mawell-dilaton theory with a non-trivial dilaton profile: 
\begin{align}
&\CL_G = R - 2 \Lambda, && \CL_\phi = -(\partial \phi)^2- V(\phi),\\ 
&\CL_A = -\frac{1}{4}Z_A(\phi) F_{A,ab}F_{A}^{\;\;ab}, && \CL_V =
-\frac{1}{4} Z_V(\phi)F_{V,ab} F_{V}^{\;\;ab},
\end{align}
having used the notation of the action in Eq. \eqref{eqn:general-action}. This is an extension of the case studied in \cite{Gursoy:2014boa}, which includes the gravitational anomaly and anomalous conductivities that follow from a response to a small vortex. 

The theory has two charges that are conserved along the radial direction at zeroth-order in the boundary-derivative expansion. The expressions follow from the $a = \mu$ component of the Maxwell's equations: 
\begin{align}
Q_5 &= r^3\sqrt{\frac{g}{f} } Z_A \partial_r A_t ,\\
Q &= r^3\sqrt{\frac{g}{f}} Z_V \partial_r V_t .
\end{align}
At first order in derivatives, the two conserved currents $\CJ^\mu_5$ and
$\CJ^\mu$ are given by
\begin{equation}
\begin{aligned}
\CJ^\mu_5 &=  \left[ Q_5 h + r^3\sqrt{f g } Z_A  \partial_r a \right] \omega^\mu
+ \CJ^\mu_{5,CS} , \\
\CJ^\mu &=  \left[ Q h  + r^3 \sqrt{f g } Z_V \partial_r v\right] \omega^\mu +
\CJ^\mu_{CS}.
\end{aligned}
\label{eqn:membrane-current-J5-J-Einstein}
\end{equation}
Thus, we can immediately read off the membrane currents:
\begin{align}
\delta J^\mu_{5,mb} &=  r^3  \sqrt{\frac{g}{f} } Z_A \partial_r a,\\
\delta J^\mu_{mb} &= r^3  \sqrt{\frac{g}{f} } Z_V  \partial_r v .
\end{align} 
Moreover, the regularity of the black hole horizon implies that that metric fluctuation has to vanish at
the horizon \cite{Gursoy:2014boa}, i.e. $h(r_h)=0$. At the horizon, the two currents $\CJ^\mu_5(r_h)$ and $\CJ^\mu(r_h)$ are therefore completely determined by the anomalous terms $\CJ^\mu_{5,CS}(r_h)$ and $\CJ^\mu_{CS}(r_h)$. 

Next, we investigate the behaviour of $\CJ^\mu_5$ and $\CJ^\mu$ at the boundary. Substituting the near-boundary solutions \eqref{eqn:near-boundary-h} into \eqref{eqn:membrane-current-J5-J-Einstein}, one can see that $Q_5 h$ and $Q h$ are sub-leading, which implies that $\CJ^\mu_5$ and $\CJ^\mu$ at $r \to \infty$ become determined by the membrane currents evaluated at the boundary.

\subsection{Four-derivative Einstein-Maxwell theory} 
\label{section:4-derivatives}

In this section, we consider the most general four-derivative theory of massless gravitons and gauge fields. The action $\mathcal{L}_A$ can be written as (see \cite{Gross:1986mw,Anninos:2008sj,Kats:2006xp,Myers:2009ij,SGNonPert}):
\begin{equation}\label{EMaction4d}
\begin{aligned}
\mathcal{L}_A =& -\frac{1}{4} F_{ab}F^{ab} + \alpha_4 R F_{ab} F^{ab} +
                  \alpha_5 R^{ab} F_{ac} F_b^{\;\;c} + \alpha_6
                  R^{abcd} F_{ab}F_{cd} +\alpha_7
                  (F_{ab}F^{ab})^2\\
& + \alpha_8 \nabla_a F_{bc} \nabla^a F^{bc} + \alpha_9\nabla_a
   F_{bc} \nabla^b F^{ac}+\alpha_{10} \nabla_a F^{ab} \nabla^c
   F_{cb}+\alpha_{11} F^{ab} F_{bc} F^{cd}F_{da},
\end{aligned}
\end{equation}
and similarly $\CL_V$. Note that in Eq. \eqref{EMaction4d}, all indices $A$ denoting that $F_{ab}$ is the axial field strength have been suppressed. The conserved current two-form, $H^{ab}_5$, in this theory is
\begin{equation}
\begin{aligned}
H^{ab}_5 =& -F^{ab} + 4 \alpha_4 RF^{ab} + 2\alpha_5 (R^{ac}F_c^{\;\;b}
-R^{bc}F_c^{\;\;a} ) + 4\alpha_6 R^{cdab}F_{cd} \\
& + 8 \alpha_7 F_{cd}F^{cd} F^{ab} - 4\alpha_8 \Box F^{ab} - 2\alpha_9 \nabla_c
(\nabla^a F^{cb} - \nabla^b F^{ca}) \\
&+ 2\alpha_{10} (\nabla^b\nabla_c F^{ca} - \nabla^a\nabla_c F^{cb}) + 8\alpha_{11} F^{bc}F_{cd}F^{da}.
\end{aligned}
\label{eqn:conserved-current-2-form-4-deriv-Maxwell}
\end{equation}
The current $\mathcal{J}^i_5$ is then 
\begin{equation}
\CJ^\mu_5=
\CJ^\mu_{5,\text{Maxwell}}+\sum_{n=4}^{11} \alpha_n \CJ^\mu_{5,(n)} + \CJ^\mu_{CS},
\end{equation}
where $\CJ^\mu_{5,\text{Maxwell}}$ is the axial current that follows from the two-derivative Maxwell action analysed in Section \ref{section:Einstein-Hilbert-Maxwell-gravity}. The remaining terms, $\CJ^\mu_{5,(n)}$, all have the schematic form
\begin{equation}
\CJ^\mu_{5,(n)} = \left[ C_{n,1} h + C_{n,2} \partial_r h + C_{n,3} \partial^2_r h + D_{n,1} \partial_r a +
D_{n,2} \partial^2_r a + D_{n,3} \partial^3_r a \right] \omega^\mu ,
\end{equation}
where the coefficients $C_{n,i}$ and $D_{n,i}$ depend on the background and
parameters of the action. The full expressions for these coefficients are lengthy and will not be presented here. 

Near the non-extremal horizon (assumed to exist), the metric must behave as in Eqs. \eqref{nearH1} and  \eqref{nearH2}. What we find is that when evaluated at the horizon, all coefficients except $C_{n,1}$ vanish. This result therefore precisely agrees with the structure of $\CJ^\mu_5$ predicted in \eqref{eqn:non-CS-CJ5-general}, which followed from our general treatment of $H^{r\mu}_5$ in Section \ref{section:any-higher-derivative}. At the horizon, the full set of $\CJ^\mu_{5,(n)}$ is given by
\begin{equation}
\begin{aligned}
\CJ^\mu_{5,(4)} (r_h) &= -\frac{2 r_h^2\sqrt{g_1}A_t'}{f_1^{3/2}} \left( 20 f_1g_1
+ 3f_2g_1 r_h + f_1g_2 r_h \right)
h(r_h) \,\omega^\mu, \\
\CJ^\mu_{5,(5)} (r_h) &= - \frac{r_h \sqrt{g_1}A_t'}{f_1^{3/2}} \left(
14 r_hf_1g_1 + 2 r_h^2 g_1 f_2 +r_h^2 f_1g_2  \right)
h(r_h)\,\omega^\mu,\\
\CJ^\mu_{5,(6)} (r_h)  &= - \frac{2  r_h^2 \sqrt{g_1} A_t'}{f_1^{3/2}} \left( 8f_1g_1
+ 3r_hg_1f_2 +r_h f_1g_2 \right) h(r_h)\, \omega^\mu,\\
\CJ^\mu_{5,(7)} (r_h)  &= -\frac{16  r_hg_1^{3/2}(A_t')^3}{f_1^{3/2}}h(r_h) \,\omega^\mu,\\
\CJ^\mu_{5,(8)}(r_h) &= - \frac{2 8r_h^3\sqrt{g_1}}{f_1^{3/2}} \left( -g_1f_2 + 
f_1g_2+2f_1g_1A_t''/A_t'
\right) h(r_h)\, \omega^\mu,\\
\CJ^\mu_{5,(9)} (r_h) &= \frac{1}{2} \CJ^\mu_{5,(8)} ,\\
\CJ^\mu_{5,(10)} (r_h)  &= \frac{ r_h^2\sqrt{g_1}}{f_1^{3/2}}\left(
 6f_1g_1 -r_h g_1f_2 + r_h f_1g_2 + 2r_h f_1g_1A_t''/A_t' \right) h(r_h) \, \omega^\mu,\\
\CJ^\mu_{5,(11)} (r_h)  &= - \frac{1}{2}\CJ^\mu_{5,(7)}.
\end{aligned}
\end{equation}
Finally, imposing the horizon Ricci scalar regularity condition (see the discussion after Eq. \eqref{eqn:non-CS-CJ5-general}), $h(r_h)=0$, we find that all $J^\mu_{5,(n)} (r_h) = 0$. 

At the AdS boundary ($r\to\infty$), we further find that all coefficients $C_{n,i} \sim r^{-m}$, where $m>0$. With this explicit verification, our results imply that the most general gauge- and diffeomorphism-invariant four-derivative theory \eqref{EMaction4d} satisfies the condition \eqref{eqn:condition-for-universality} and that the anomalous conductivities in its dual all have the universal form of Eq. \eqref{eqn:universal-anomalous-conductivity-electric}.

\subsection{Theories without horizons and theories with scaling geometries at zero temperature}
\label{section:counter-examples}

In this section, we consider two classes of backgrounds, each one a possible solution of the Einstein-Maxwell-dilaton theory of Section \ref{section:Einstein-Hilbert-Maxwell-gravity}. The first one belongs to the family of soft/hard wall model that are dual to a field theory with a mass gap \cite{Csaki:1998qr,Karch:2006pv,Gursoy:2007er,Batell:2008zm}. The second example is the scaling
geometry that can arise as a solution of the Einstein-Maxwell-dilaton theory at zero temperature (see e.g. \cite{Charmousis:2010zz}). What we show is that the criterion for the universality of anomalous conductivities, i.e. Eq. \eqref{eqn:condition-for-universality}, is still satisfied in the gapped system. However, the conductivities can no longer computed by using the replacement rule in the form stated in Eq. \eqref{eqn:replacement-rule-conductivity}. For the scaling geometries, the universality may be violated due to the presence of naked singularities. A way to retain a holographic theory at zero temperature in which the condition \eqref{eqn:condition-for-universality} is satisfied is to put very strong constraints on the geometry that avoid the naked singularity. These constraints restrict the allowed range of value of the hyperscaling violation exponent, $\theta$, and the dynamical critical exponent, $z$. 

Let us start with an example of the soft/hard wall geometry at zero density. In an un-boosted frame, the metric for these models can be written as
\begin{equation}
ds^2 = e^{-(M/u)^\nu} \left( -u^2 d\bar{t}^2 + \frac{du^2}{u^2} + u^2 (d\bar x^2+d\bar
  y^2+d\bar z^2)  \right),
\end{equation}
where the parameter $M$ sets the scale of the mass gap. The nature of the spectrum is also controlled by the parameter $\nu$: while the gapped spectrum is continuous above the gap when $\nu =1$, it is discrete when $\nu >1$. The hard wall model
in which the AdS radius is capped off at $u\ll M$ corresponds to the limiting value of $\nu
\to \infty$ \cite{Gursoy:2007er,Batell:2008me}. 

One can change coordinates of the above metric to bring them to the form of
\eqref{eqn:background-solution-unboosted} by redefining the radial coordinate as $r =
e^{-\frac{1}{2}(M/u)^\nu} u$. In the deep IR region, $u\ll M$, the functions $f(r)$ and $g(r)$ can be written as 
\begin{align}
f_{IR}(r) = 1,&& g_{IR}(r) = g\left(u\ll M\right)=\nu^2 \left(\frac{M}{u} \right)^\nu e^{M/u}.
\label{eqn:wall-f-and-g}
\end{align}
Despite there being no horizon, the dual of the above geometry can still have non-zero temperature; it can be interpreted as a thermal state before undergoing a phase transition to the black hole phase at high temperature, analogously to the Hawking-Page transition \cite{Herzog:2006ra}. 

The two currents, $\CJ^\mu_5$ and $\CJ^\mu$, must now be evaluated at $r = 0$ and at the boundary $(r = \infty)$. Because the geometry is still asymptotically AdS, their near-boundary behaviour is the same as in all the cases studied before. The fact that $g(r)$ exponentially diverges in the IR appears problematic at first. However, the volume form, which is proportional to $\sqrt{-G}$, is exponentially suppressed. Evaluating $\CJ^i_5$ at $u=0$, one finds that $\CJ^{\mu}_{5,mb}(0) + \CJ^{\mu}_{5,r}(0)  = 0$ as in \ref{section:any-higher-derivative}. Thus, the universality condition \eqref{eqn:condition-for-universality} is still satisfied. 

On the other hand, the Chern-Simons current $\CJ^\mu_{5,CS}$ no longer behaves the same way. Although the profiles of the gauge fields $A_t$, and $V_t$ can be assumed to asymptote to a constant value at $r = 0$, the derivative of $f'$ can no longer be interpreted as the temperature of the dual theory (substituting \eqref{eqn:wall-f-and-g} into
\eqref{eqn:membrane-and-CS-current}, we see that $\CJ^\mu_{5,CS}$ has no temperature
dependence). Therefore, in the confining phase, the replacement rules discussed in the Appendix \ref{app:anomP} are no longer applicable even if the condition \eqref{eqn:condition-for-universality} is satisfied. The above statements
also apply to the AdS soliton-like geometries. 

Next, we explore the scaling geometries at zero temperature. In the un-boosted frame, the metric can now be written as 
\begin{equation}
ds^2 = r^2 \left( -r^{n_0} dt^2 +d\bar x^2+d\bar y^2+d\bar z^2   \right) + \frac{dr^2}{r^{n_1}},
\end{equation}
or in terms of $\theta$ and $z$,
\begin{align}
n_0 = 2 + \frac{6(z-1)}{3-\theta},\qquad n_1 = 2+ \frac{2\theta}{3-\theta}.
\end{align}
As mentioned in \cite{Charmousis:2010zz}, many of these geometries contain a naked singularity. As a result, the scalars $\{ \CS_{A,n},\CS_{G,n} \}$ used in Eq. \eqref{eqn:decomposed-CC} no longer have to be finite. Such systems can therefore easily violate the universality condition \eqref{eqn:condition-for-universality}. Thus, the universality of the anomalous conductivities is no longer guaranteed in the presence of a naked singularity. In this work, we do not study in detail what happens to anomalous conductivities in such cases and whether they nevertheless remain universal for some geometries.

Are there special values of $z$ and $\theta$ for which it is easy to see that the condition \eqref{eqn:condition-for-universality} remains satisfied? In other words, what are the ranges of $\{ z,\theta\}$ for which the theory has no naked singularity? This problem was addressed in \cite{Copsey:2012gw}, where it was found that the geometries that satisfy either one of the following two conditions,
\begin{equation}
n_0 = n_1 =2 ,\qquad n_0 = n_1 \ge 4,
\label{eqn:condition-no-naked-singularity}
\end{equation}
have no naked singularities. The authors assumed that the matter content has to satisfy the null energy condition, which, for this geometry, is equivalent to imposing the following two inequalities: 
\begin{equation}
n_0 \ge n_1, \qquad (n_0-2)(n_0+n_1+4) \ge 0.
\label{eqn:NEC}
\end{equation}
The first solution in \eqref{eqn:condition-no-naked-singularity} is simply the empty AdS solution with $z=1$ and $\theta=0$. The second solution (or a family of solutions) is more involved and requires non-trivial matter to support such geometries. 

Of particular interest are charged dilatonic black holes with $z \to \infty$, $\theta \to -\infty$ and a fixed ratio $-\theta/z=\eta$, dual to strongly interacting theories with finite density (see e.g. \cite{Gubser:2009qt,Huijse:2011ef}). While such systems still satisfy the null energy condition, the geometries nevertheless exhibit a naked singularity at zero temperature. This means that unless there is a way to resolve the singularity, the universal structure of anomalous conductivities, although not necessarily, may be violated at zero temperature for all values of $\eta$. One way to resolve this issue, as mentioned in \cite{Gubser:2009qt} for $\eta=1$, is to lift the black hole solution to a ten- or eleven- dimensional solution of string or M-theory \cite{Cvetic:1999xp}. To study such solutions, one also needs to find the ten- and eleven-dimensional analogues of the Chern-Simons terms ($\CL_{CS}$ in \eqref{eqn:LA-LV-LCS}). In case of a supergravity setup, this was studied in \cite{Klebanov:2002gr} and many subsequent works. An explicit computation of chiral magnetic conductivity, $\sigma_{JB}$, in a top-down setup of probe flavour branes can be found in \cite{Hoyos:2011us}. More generally, it is plausible that the problems of IR singularities can be avoided when they are of the ``good type" \cite{Gubser:2000nd}.\footnote{We thank Umut G\"{u}rsoy for discussions on this point.} In such scenarios, it may be the case that so long as the naked singularity can be cloaked by an infinitesimal horizon, the existence of universality can be extended to very small temperatures. What is clear is that at strictly zero temperature, the regularity of (small) metric perturbations is no longer well-defined. We defer a more detailed study of these issues and of top-down constructions to future works.

\subsection{Bulk theories with massive vector fields}
\label{section:massive-vector-fields}

In this section, we comment on the universality of anomalous conductivities in field theories with mixed, gauge-global anomalies. Such theories exhibit the following anomalous Ward identity:
\begin{equation}
\partial_\mu \left\la J^\mu_5\right\ra = \beta \epsilon^{\mu\nu\rho\sigma}   \CF_{\mu\nu} \CF_{\rho\sigma} +
\left(\text{global anomaly terms}\right) ,
\label{eqn:anomalous-Ward-identitiy-gluon}
\end{equation}
where $\CF_{\mu\nu}$ is the field strength of the gluon fields (e.g. in QCD). The global anomaly terms were stated in Eq. \eqref{eqn:non-conserved-current}. As shown by perturbative quantum field theory calculations \cite{Neiman:2010zi,Golkar:2012kb,Hou:2012xg,Jensen:2013vta}, the anomalous conductivities in such theories are known to be renormalised, i.e. they receive quantum corrections.

Holographic models dual to theories with the anomalous Ward identity of the form of Eq. \eqref{eqn:anomalous-Ward-identitiy-gluon} were proposed and studied in \cite{Klebanov:2002gr,Casero:2007ae,Jimenez-Alba:2014iia,Gursoy:2014ela,Jimenez-Alba:2015awa}. In this work, we focus on the bottom-up construction of \cite{Jimenez-Alba:2014iia}, where the following terms are added to the bulk action \eqref{eqn:general-action}:
\begin{equation}
\Delta S = \int d^5x \sqrt{-G} \left( - \frac{m^2}{2} (A_a - \partial_a
  \theta)(A^a - \partial^a \theta) -\frac{\kappa}{3} \epsilon^{abcde} (\partial_a \theta) F_{bc}F_{de}\right)
\end{equation}
We have set the vector and the gravitational Chern-Simons terms to zero, i.e. $\gamma = \lambda = 0$ (see Eq. \eqref{eqn:non-conserved-current}). The scalar field $\theta$ is
the St\"uckelberg axion.

A holographic theory with $\Delta S$ in the action can clearly evade the arguments of the proof of universality from Section \ref{section:anomaly-induced-electric-current}. The reason is that the equation of motion for a massive vector field cannot be written in the form of Eq. \eqref{eqn:Maxwell-eoms}. The right-hand-side of \eqref{eqn:Maxwell-eoms} now contains terms which explicitly depend on $A_a$ and one cannot reduce the equations into a total derivative form, $\partial_r \CJ^\mu = 0$. Hence, in models with massive vector fields, dual to field theories with mixed, gauge-global anomalies, anomalous conductivities can be renormalised. This is consistent with field theory calculations mentioned above. More precisely, from the point of view of field theory, the operators associated with anomalous transport are renormalised along the renormalisation group flow. In gravity, they depend on the entire bulk geometry and thus the condition of horizon regularity is not sufficient to ensure universality. In relation to our discussion about universality in field theory (see the Introduction \ref{sec:Intro}), it would be interesting to understand what precisely happens to the arguments of the regularity of one-point functions on a cone in such cases.

\section{Discussion}
\label{section:discussion}

In this work, we studied the coupling constant dependence of the universality of chiral conductivities associated with the anomalous axial and vector currents in holographic models with arbitrary higher-derivative actions of the metric, gauge fields and scalars. We showed that so long as the action (excluding the Chern-Simons terms) was gauge- and diffeomorphism-invariant, the membrane paradigm construction for the chiral conductivities remained valid, resulting in universal chiral conductivities (see Eq. \eqref{eqn:universal-anomalous-conductivity-electric}). The proof assumed the existence of a regular, non-extremal black brane with an asymptotically AdS geometry. This result is valid for an infinite-order expansion of coupling constant corrections to holographic results at infinite coupling. Hence, it is complementary to perturbative field theory proofs (expanded around zero coupling) of the non-renormalisation of chiral conductivities in systems with global anomalies and therefore of the anomalous Ward identities with the form of Eq. \eqref{eqn:non-conserved-current}. Furthermore, our paper also explored cases which may violate universality, in particular, in cases with naked singularities and massive vector fields that explicitly violate Eq. \eqref{eqn:non-conserved-current} through mixed, gauge-global anomalies.

This work provides a consistency test of holography in its ability to reproduce the expected non-renormalisation of global Ward identities at the level of (non-zero temperature and density) transport in very general bulk constructions that include arbitrary higher-derivative actions. Furthermore, we believe that the methods presented in this work can be of wider use to other holographic statements of universality that employ the membrane paradigm.

An important conceptual question that remains is the precise relation between the regularity condition of our constructions at the horizon and properties of their dual field theories. It is tempting to speculate that the regularity of the background geometry is related to the regularity of one-point functions on a cone that fix $\tilde c$ and ensure universality of anomalous conductivities in field theory (see discussion after Eq. \eqref{s2}).\footnote{We thank the anonymous JHEP referee for a discussion regarding this point.}

We end this paper by listing some problems that are left to future works. Most importantly, there exists another anomalous conductivity in the stress-energy tensor, which can be sourced by a small vortex, $\delta T^{\mu\nu} = \sigma^\epsilon u^{(\mu} \omega^{\nu)}$. The analysis of this conductivity was not performed in this work. In the fluid-gravity framework, $\sigma^\epsilon$ was studied in the Einstein-Maxwell theory by \cite{Azeyanagi:2013xea}. Forming a conserved bulk current for computing components of the stress-energy tensor tends to be significantly more complicated than for those of a boundary current. However, it may be possible to achieve this by using the Hamiltonian methods recently employed for the calculations of the thermo-electric DC conductivities \cite{Donos:2014cya,Donos:2015gia,Banks:2015wha} in two-derivative theories, which should be extended to computations of anomalous transport in higher-derivative theories. 

One may also wonder what happens to anomalous transport in inhomogeneous and anisotropic systems. In standard non-anomalous transport, it is known that universal relations can be violated, e.g. in $\eta / s$ \cite{Mamo:2012sy,Jain:2014vka,Jain:2015txa,Hartnoll:2016tri,Alberte:2016xja,Burikham:2016roo}. While analysing such systems is in general significantly more difficult, the existence of the membrane paradigm, as e.g. in case of the DC thermo-electric conductivities \cite{Donos:2015gia,Banks:2015wha,Donos:2015bxe}, may still enable one to prove general statements about the behaviour of conductivities in disordered systems \cite{Grozdanov:2015qia,Grozdanov:2015djs}. These methods remain to be explored in the context of anomalous transport.

In even-dimensional theories, anomalous conductivities are directly related to the parity-odd hydrodynamic constitutive relation of \cite{Son:2009tf,Landsteiner:2011iq,Jensen:2012jy}. These parity-odd terms are related to global anomalies. In odd dimension, one can still construct hydrodynamics with parity-odd terms, as e.g. in \cite{Jensen:2011xb}. A well-known parity-odd transport coefficients is the Hall viscosity \cite{Avron:1995fg,Avron:1997}. This quantity has relations to topological states of matter, such as fractional quantum Hall systems (see e.g. \cite{Hoyos:2014pba} and references therein). A holographic theory with non-zero Hall viscosity can be obtained by adding a topological term similar to the dimensionally-reduced gravitational Chern-Simons term \cite{Saremi:2011ab}. Recently, in \cite{Haehl:2015pja}, the constitutive relation term associated with the Hall viscosity was generalised to a class of hydrodynamic terms that resemble the Berry curvature. Despite these similarities, there is no known non-renormalisation theorem for parity-odd transport coefficients in odd dimensions. 

Lastly, we point out that many recent works have found novel structures in entanglement entropy of theories with anomalies
\cite{Castro:2014tta,Iqbal:2015vka,Nishioka:2015uka,Nishioka:2015uka,Azeyanagi:2015uoa,Belin:2015jpa}. As a result of non-renormalisation, one may expect there to exist strong constraints on the structure of extremal bulk surfaces associated with entanglement entropy. It would be interesting to better understand the connection between geometric constraints on holographic entanglement entropy and non-renormalisation theorems for anomalies.

\acknowledgments{The authors would like to thank Richard Davison, Misha Goykhman, Nabil
  Iqbal, Aron Jansen, Janos Pol\'{o}nyi, Petter S\"aterskog, Koenraad Schalm and Jan Zaanen for stimulating discussions. We are
  particularly grateful to Umut G\"{u}rsoy and Javier Tarr\' io for their comments on
  the draft of the manuscript. S. G. is supported in part by a VICI grant of
  the Netherlands Organisation for Scientific Research (NWO), and by the Netherlands
  Organisation for Scientific Research/Ministry of Science and Education (NWO/OCW). The
  work of N. P. is supported by the DPST scholarship from the Thai government and by Leiden University. }

\appendix
\section{Anomaly polynomials and the replacement rule}\label{app:anomP}

As noted in the Introduction, the full set of chiral conductivities \eqref{CondDef} can be encoded in the anomaly polynomial defined in terms of the Chern-Simons action \cite{Harvey:2005it,Bilal:2008qx,Loganayagam:2012zg,Jensen:2013rga}: 
\begin{align}
\CP(F,R) =  d S_{CS} \left[A,\Gamma\right] .
\end{align}
If we restrict ourselves only to global anomalies in four spacetime dimensions, then the anomaly polynomial can be written as
\begin{equation}
\CP = \frac{\kappa}{3} \left( {F}_A \wedge {F}_A \wedge {F}_A\right) + \gamma \left( {F}_A \wedge {F}_V
  \wedge {F}_V\right) + \lambda \left( {F}_A \wedge {R}^\mu_{\;\;\nu} \wedge
  {R}^\nu_{\;\;\mu} \right).
\label{eqn:anomaly-polynomial}
\end{equation} 
The replacement rule states that, for an anomaly polynomial $\CP$, one can define the
generating function $\CG[\mu_5,\mu,T]$: 
\begin{equation}
\CG \left[\mu_5,\mu,T \right] = \CP \left[{F}_A \to \mu_5,\, {F}_V \to \mu, \,\tr \,  R^{2} \to 2 (2\pi T)^2\right] ,
\label{eqn:def-replacement-rule}
\end{equation}
where $T$ is the temperature and $\mu_5,$ and $\mu$ are chemical potentials associated with the axial and the vector currents
$J^\mu_5$ and $J^\mu$. The anomalous conductivities can then be computed by using
\begin{align}
&\sigma_{J_5 B} = -\frac{\d^2 \CG}{\d\mu_5\d \mu}, && \sigma_{J B} = -\frac{\d^2 \CG}{\d \mu \d\mu},\nn
&\sigma_{J_5 \omega} = \frac{\d \CG}{\d \mu_5 }, && \sigma_{J\omega} = \frac{\d \CG}{\d\mu}.
\label{eqn:replacement-rule-conductivity}
\end{align}
For the anomaly polynomial in \eqref{eqn:anomaly-polynomial}, the anomalous conductivities are precisely those stated in Eq. \eqref{eqn:universal-anomalous-conductivity-electric}. 

In the work of \cite{Azeyanagi:2013xea}, the replacement rule \eqref{eqn:def-replacement-rule} with \eqref{eqn:replacement-rule-conductivity} was derived for a field theory dual to the AdS Reissner-N\" ordstrom background. Our work can be seen a check of the validity of this replacement rule prescription for more general, higher-derivative holographic theories.

\bibliography{biblio}
\end{document}